\documentclass[a4paper,11pt,preprintnumbers]{article}

\usepackage{jheppub} 
\usepackage{amsthm}
\usepackage{multirow,bm}

\usepackage[T1]{fontenc} 

\title{\boldmath Measurement of physical parameters with a weight function method and its application to the Higgs boson mass reconstruction}

\author[a,1]{S. Kawabata,\note{Corresponding author.}}
\author[a]{Y. Shimizu,}
\author[a]{Y. Sumino}
\author[b]{and H. Yokoya}

\affiliation[a]{Department of Physics, Tohoku University,\\Sendai 980-8578, Japan}
\affiliation[b]{Department of Physics, University of Toyama,\\Toyama 930-8555, Japan}

\emailAdd{skawabata@tuhep.phys.tohoku.ac.jp}
\emailAdd{shimizu@tuhep.phys.tohoku.ac.jp}
\emailAdd{sumino@tuhep.phys.tohoku.ac.jp}
\emailAdd{hyokoya@sci.u-toyama.ac.jp}

\abstract{We propose a new method to measure various physical parameters, using characteristic weight functions. The method requires only lepton energy distribution and ideally it does not depend on the velocity of the parent particle. We demonstrate an application of this method by simulating a reconstruction of the Higgs boson mass in the $H\rightarrow WW \rightarrow l\nu l\nu$ decay mode at the LHC. We show that systematic errors are suppressed compared to statistical errors. In the vector boson fusion channel, the statistical accuracy of the mass determination is estimated to be $+12\%$ and $-14\%$ at an integrated luminosity of 100\,fb$^{-1}$, assuming the Higgs boson mass to be 125\,GeV and $\sqrt{s}=14$\,TeV.}

\begin{document} 

\begin{flushright}
TU--939\\
UT-HET-082
\end{flushright}
\vspace*{-15mm}

\maketitle
\flushbottom

\section{Introduction}
\label{sec:intro}

A search for the Higgs boson and new physics beyond the Standard Model has been intensively performed at the LHC. Recently the ATLAS and CMS collaborations announced the presence of a new particle which is compatible with the Standard Model Higgs boson~\cite{:2012gk,:2012gu}. It is expected that the study of the Higgs boson will elucidate the mechanism of the electroweak symmetry breaking and the origins of particle masses.

Once we find the Higgs boson or other new particles, the next step is to understand their properties in detail. It is known, however, that accurate measurements at the LHC are generally subject to the following difficulties:
\begin{itemize}
	\item In many cases of interest there are undetected particles in the final state, hence it is difficult to reconstruct their missing momenta accurately. For instance, in such cases it becomes non-trivial to apply the most typical method to measure the mass of a parent particle, namely, to reconstruct the invariant mass of all the momenta in the final state. 
	\item Since jets overlap with one another and each jet contains many neutral particles, it is difficult to measure the momentum of each jet accurately compared to that of a charged lepton $e$ or $\mu$.
	\item The center-of-mass system of the collision at parton level is not known. Moreover, the parton distribution function (PDF), which is used for calculations of cross sections and kinematical distributions, has relatively large uncertainties. Therefore, predicting or constraining kinematics of events from the initial states involves uncertainties inevitably.
\end{itemize}

To overcome these difficulties, many methods for reconstruction of kinematics have been proposed and developed~\cite{Barr:2010zj}. Some of them utilize a kinematic variable $m_{T2}$~\cite{Lester:1999tx, Barr:2003rg} or MAOS momenta~\cite{Cho:2008tj}, which have the advantage that they can be applied to processes with missing particles. However, since these methods make direct use of missing momenta, they are accompanied by systematic uncertainties on jet momenta through missing momenta. Since these uncertainties are relatively large, these methods are not adapted to accurate measurements but rather for discovery physics. 

In this paper, we propose a new method for determination of various physical parameters, which can suppress systematic uncertainties originating from the above factors. The method has the following features:
\begin{itemize}
	\item Only lepton energy distribution is needed as input measured quantities. We can apply this method to processes which contain at least one lepton in their final states. It does not matter if any jets or missing particles are included in the final states.
	\item In ideal cases, this method does not depend on the velocity of the parent particle. Since it is difficult to know the velocity of the parent particle accurately from its production process, due to uncertainties in the PDF, ISR, etc., with this method we can avoid these uncertainties.
\end{itemize}
This method is valid if the parent particle is scalar or unpolarized. We can determine any parameters which enter the lepton energy distribution in the rest frame of the parent particle, except for parameters which only affect the normalization of the distribution. 

The method with the above features is realized by using a weight function $W$ with the following characteristics. We integrate the lepton energy distribution in the laboratory frame $D(E_l)$ weighted by a function $W(E_l,\lambda)$, and write the weighted integral as $I(\lambda)$:
\begin{equation}
	I(\lambda) ~\equiv~ \int dE_l \,D(E_l) W(E_l\,,\lambda)\,,
\end{equation}
where $\lambda$ is a physical parameter to be measured. A typical example of $\lambda$ is the mass of the parent particle. Then $I(\lambda)$ satisfies
\begin{itemize}
	\item[i)] ~$I(\lambda)~=~0$ ~if $\lambda$ takes the true value.
	\item[ii)] Although $D(E_l)$ depends on the velocity distribution of the parent particle, the property i) holds true for any of those distributions. 
\end{itemize}
That is, $I(\lambda)$ can judge whether $\lambda$ is the true value or not, independent of velocity distribution of the parent particle. There exist an infinite number of such characteristic weight functions for a given $\lambda$, and we can construct them explicitly.

A lepton energy distribution which we can obtain from an experiment is affected by detector effects and event selection cuts, and furthermore includes backgrounds. However, as we will demonstrate later, these effects can be estimated with small systematic uncertainties by Monte Carlo simulations.

We apply the above method to a reconstruction of the Higgs boson mass at the LHC. The mass of the Higgs boson has been measured at the LHC as $m_H=125.7\pm0.3\pm0.3$\,GeV by the CMS collaboration and $m_H=125.5\pm0.2\,\,^{+0.5}_{-0.6}$\,GeV by the ATLAS collaboration, using the $H\rightarrow \gamma \gamma$ and $H\rightarrow ZZ$ modes~\cite{CMS:2013Morio, ATLAS:2013Comb}. In order to reveal properties of the Higgs boson more closely, it is important to analyze further various production or decay channels. The $H\rightarrow WW$ mode is one of the most important decay modes with the second largest branching ratio for $m_H \approx 125$ GeV, associated with charged leptons with relatively large $p_T$ in the final state if $W$ bosons decay leptonically. In this mode, however, because of the missing momenta in the final state, we cannot reconstruct the invariant mass of the Higgs boson. For this reason, it is difficult to use this mode for a determination of the Higgs boson mass. The mass measurement using the $WW$ decay mode has not been performed in the LHC experiments so far. There are some methods proposed by theorists to reconstruct the Higgs boson mass using the $WW$ mode. The methods proposed in refs.~\cite{Barr:2009mx, Choi:2009hn, Choi:2010wa} utilize variables sensitive to $m_H$. However, they use missing momenta directly, which would cause sizable systematic uncertainties. Another method utilizes the cross section and lepton $p_T$ spectra~\cite{Davatz:2006ja}, which would be affected by uncertainties of the PDF of initial partons. In this paper, we investigate a mass reconstruction of the Higgs boson in the $WW$ mode as an application of the weight function method. We perform simulation analysis for a reconstruction of the Higgs boson mass in the $H\rightarrow WW \rightarrow l\nu l\nu$ (with $l=e, \mu$) decay mode.

The major part of our new method has been reported briefly in a letter article~\cite{Kawabata:2011gz}. There, we applied our method to a reconstruction of the Higgs boson mass. At that time, however, the mass of the Higgs particle was much more obscure than it is today, and we performed a MC simulation analysis assuming $m_H=150$\,GeV. We confirmed that indeed systematic uncertainties are suppressed and under control in our method. Moreover, since the statistical error has been estimated to be $2\%$ corresponding to an integrated luminosity of 100\,fb$^{-1}$ for this mass value, the suppression of systematic uncertainties could be practically important. The purpose of the present paper is two fold. First, we give a detailed description of our method, with details of the theoretical formulas, elucidating various characteristics of our method. Secondly, we perform a MC simulation analysis using today's realistic Higgs boson mass value. It is well known that with this mass value a Higgs mass reconstruction using the $H \rightarrow WW \rightarrow l\nu l\nu$ mode is quite challenging, due to limited statistics and large backgrounds. So far, we find no detailed studies in the literature which assume $m_H \approx 125$\,GeV and use this decay mode. Thus, we consider it worth presenting a study for this case, including estimates for both statistical and systematic uncertainties, in order to provide a reference point.

After our first analysis~\cite{Kawabata:2011gz}, a similar method was proposed in ref.~\cite{Agashe:2012bn}. This method uses the fact that in the two-body decay of an unpolarized particle, the peak in the energy distribution of the massless daughter particle is invariant under the Lorentz boost, and equal to its (monochromatic) energy in the mother particle's rest-frame. Although this simple method works too for the two-body decay, our method utilizes more general features of the energy distribution of boosted particles and can be applied to decays into any number of particles.

The paper is organized as follows. In section~\ref{sec:WeightFunction}, we prove the existence of characteristic weight functions, and propose a new method to determine physical parameters using these weight functions. In section~\ref{sec:HiggsSimulation}, we apply this method to a reconstruction of the Higgs boson mass in the $H \rightarrow WW \rightarrow l\nu l\nu$ mode at the LHC, performing a simulation analysis. We estimate the accuracy of the mass determination and examine the properties of the weight function method. In section~\ref{sec:Conclusion}, we summarize our analysis. Details of the computation of the lepton energy distribution are given in the Appendix.

\section{Characteristic weight functions}
\label{sec:WeightFunction}

In this section, we consider a decay of a scalar or unpolarized particle into many bodies including at least one lepton ($e$ or $\mu$) in the final state. For such a decay there exist weight functions $W(E_l)$ which have the following characteristics.
\begin{itemize}
	\item[i)] The integral of the lepton energy distribution in the laboratory frame $D(E_l)$ weighted by $W(E_l)$ is equal to zero:
	\begin{equation}
		\int dE_l \,D(E_l) \,W(E_l) ~=~0 \,.
	\end{equation}
	\item[ii)] The property i) holds true irrespective of the velocity distribution of the parent particle.
\end{itemize} 
We prove that there exist an infinite number of such characteristic weight functions, and we present their explicit forms. Since the shape of the lepton energy distribution $D(E_l)$ depends strongly on the velocity distribution of the parent particle, the existence of the above characteristic weight functions is nontrivial.

We first discuss the case of two-body decay (section~\ref{sec:2BodyDecay}), and then generalize the argument to the case of many-body decay (section~\ref{sec:ManyBodyDecay}). Using the characteristic weight functions, we propose a method to reconstruct physical parameters (section~\ref{sec:WFM}).

\subsection{The case of 2-body decay}
\label{sec:2BodyDecay}

Suppose a parent particle $X$ decays into two particles $l$ and $Y$\,(\,$X \rightarrow l \,+\, Y$\,), where particles $X$ and $l$ satisfy the following conditions:
\begin{itemize}
	\item $l$ is massless and its energy distribution is accurately measurable.
	\item $X$ is scalar or unpolarized.
\end{itemize}

In the rest frame of $X$, the normalized energy distribution of $l$ is given by
\begin{equation}
	\mathcal{D}_0(E_l) ~=~ \delta (E_l - E_0)\,,
	\label{eq:LepEneDist_RestFrame_2BD}
\end{equation}
with
\begin{equation}
	E_0 ~\equiv~ \frac{m_X}{2} \left( 1-\frac{m_Y^2}{m_X^2}\right)\,,
\end{equation}
where $E_l$ is the energy of $l$ and $m_i$ is the mass of particle $i$. Eq.~\eqref{eq:LepEneDist_RestFrame_2BD} shows that in the rest frame of $X$ the energy of $l$ is determined uniquely.

Let us consider the same decay in a boosted frame in which $X$ has a velocity $\beta$. The phase space of 2-body decay is given by
\begin{equation}
	d\Phi_2 ~=~ \frac{1}{8\pi} ~\frac{1}{\gamma \beta m_X} 
			~\theta \left(\,\gamma \,(1- \beta)E_0 \,\leq\, E_l \,\leq\, \gamma \,(1+\beta)E_0\,\right) dE_l\,,
\end{equation}
where the step function is defined by
\begin{equation}
	\theta (~condition ~) ~\equiv~\left\{ 
		\begin{array}{ll}
			~1 & ~~~~~\mbox{if $condition$ is satisfied\,,}\\
			~0 & ~~~~~\mbox{otherwise\,.}
		\end{array}\right.
\end{equation}
Using the rapidity $y$ of $X$ in the direction of its motion, defined by
\begin{equation}
	e^{2y} ~\equiv~ \frac{1+ \beta}{1- \beta} \,\,,
\end{equation}
the normalized energy distribution in the boosted frame is expressed as
\begin{eqnarray}
	\mathcal{D} (E_l\,;\beta) ~=~ \frac{1}{2E_0 \sinh y} ~\theta \left( e^{-y} E_0 \,\leq\, E_l \,\leq\, e^y E_0 \right)\,.
	\label{eq:LepEneDist_2BodyDecay}
\end{eqnarray}

We construct a weight function $W(E_l)$ such that the integral of $\mathcal{D}(E_l\,;\beta)$ weighted by $W(E_l)$ becomes independent of the parent particle's velocity $\beta$. For convenience, we write
\begin{equation}
	W(E_l) ~=~ \left.\frac{dG(x)}{dx}\right|_{x=E_l/E_0}\,.
	\label{eq:WeightFunc_2BodyDecay}
\end{equation}
Then the weighted integral can be written as
\begin{eqnarray}
	\int dE_l \,\mathcal{D}(E_l\,;\beta) \,W(E_l) 
		~=~\, \frac{1}{2 \sinh y} \left[ \,G(e^y) -G(e^{-y})\,\right] \,.
	\label{eq:WeightedIntegral_2BodyDecay}
\end{eqnarray}
Since the right-hand side is an even part of $G(e^y)/\sinh y$,
\begin{equation}
	\frac{G(e^y)}{\sinh y} ~=~ (~\text{odd function of $y$}~) ~+~\text{const.}
	\label{eq:ConditionOnG}
\end{equation}
should be satisfied in order that the weighted integral is independent of $\beta$ (i.e., independent of $y$). Hence, the condition for $G(e^y)$ is
\begin{equation}
	G(e^y) ~=~ (~\text{even function of $y$}~) ~+~\text{const.} \times \sinh y \,.
	\label{eq:G(e^y)}
\end{equation}
Consequently, $W(E_l)$ is given by
\begin{equation}
	W(E_l) ~=~ \left.e^{-\rho} \,\left[ \,\text{(~odd function of $\rho$~)}
		~+~\text{const.}\times \cosh \rho~\right]\,\right|_{e^{\rho}=E_l/E_0} \,.
	\label{eq:WeightFunc_Odd+Cosh}
\end{equation}

In eq.~\eqref{eq:WeightFunc_Odd+Cosh}, one finds that the second term in the square bracket proportional to $\cosh \rho$ does not lead to an independent relation. In fact, if we choose $[\,\cdots]=\cosh \rho$ in eq.~\eqref{eq:WeightFunc_Odd+Cosh},
\begin{equation}
	W(E_l) ~=~ \left.\frac{\cosh\rho}{e^\rho}\right|_{e^\rho=E_l/E_0} 
		~=~ \frac{E_0^2}{2} \left(\frac{1}{E_0^2}+\frac{1}{E_l^2}\right) \,,
\end{equation}
the weighted integral satisfies
\begin{equation}
	\int dE_l \,\mathcal{D}(E_l\,;\beta)\,W(E_l) 
		~=~ \frac{E_0^2}{2} \int dE_l \,\mathcal{D}(E_l)\left( \frac{1}{E_0^2}+\frac{1}{E_l^2} \right)
		~=~ 1 \,,
	\label{eq:RelationCosh}
\end{equation}
which corresponds to $G(e^y)/\sinh y=1$ in eq.~\eqref{eq:ConditionOnG}. On the other hand, if we choose $[\,\cdots]=\sinh\rho$,
\begin{equation}
	W(E_l) ~=~ \left.\frac{\sinh\rho}{e^\rho}\right|_{e^\rho=E_l/E_0} 
		~=~ \frac{E_0^2}{2} \left(\frac{1}{E_0^2}-\frac{1}{E_l^2}\right) \,,
\end{equation}
the weighted integral satisfies
\begin{equation}
	\int dE_l \,\mathcal{D}(E_l\,;\beta)\,W(E_l) 
		~=~ \frac{E_0^2}{2} \int dE_l \,\mathcal{D}(E_l)\left( \frac{1}{E_0^2}-\frac{1}{E_l^2} \right)
		~=~ 0 \,.
	\label{eq:RelationSinh}
\end{equation}
Using the normalization condition $\int dE_l \,\mathcal{D}(E_l\,;\beta)=1$, one sees that these two relations \eqref{eq:RelationCosh} and \eqref{eq:RelationSinh} are equivalent. Therefore, the second term in eq.~\eqref{eq:WeightFunc_Odd+Cosh} is essentially included in the first term, so that we omit it hereafter.

As a result, we obtain the weight functions
\begin{equation}
	W(E_l) ~=~ \left.e^{-\rho} \,\,(~\text{odd function of }\rho~)\right|_{e^\rho=E_l/E_0}
	\label{eq:WF_2BD}
\end{equation}
which satisfy
\begin{equation}
	\int dE_l \,\mathcal{D}(E_l\,;\beta)\,W(E_l) ~=~ 0\,.
	\label{eq:IntegralZero}
\end{equation}

\begin{figure}[t]
	\begin{center}
		\includegraphics[scale=.5]{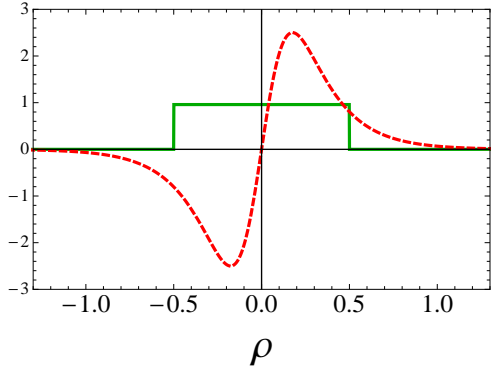}
		\caption{\label{fig:2BodyDecay_RhoDist} The normalized energy distribution $\mathcal{D}(E_l;\,\beta)$ (the solid line) and $e^{\,\rho}W(E_l)$ (the dashed line) in terms of $\rho=\log (E_l/E_0)$. $\beta$ is chosen such that $y=0.5$\,, which is the edge point of $\mathcal{D}(E_l;\,\beta)$, and $E_0$ is taken to be $1$. The odd function of $\rho$ in eq.~\eqref{eq:WF_2BD} is chosen as $5\tanh (5\rho)/\!\cosh (5\rho)$.}
	\end{center}
\end{figure}

To see how the weighted integral becomes independent of $\beta$, let us transform $E_l$ to $\rho=\log (E_l/E_0)$. Using eqs.~\eqref{eq:LepEneDist_2BodyDecay} and \eqref{eq:WF_2BD}, one finds that eq.~\eqref{eq:IntegralZero} can be expressed with $\rho$ as
\begin{equation}
	\int dE_l \,\mathcal{D}(E_l\,;\beta) \,W(E_l) 
		~=~ \frac{1}{2 \sinh y} \,\int d\rho~\theta \left( -y \,\leq\, \rho \,\leq\, y \right)\times \text{(~odd function of $\rho$~)} ~=~ 0 \,.
\end{equation}
Note that the energy distribution $\mathcal{D}(E_l\,;\beta)$ is proportional to $\theta \left( -y \,\leq\, \rho \,\leq\, y \right)$, that is, even function of $\rho$, while the weight function is proportional to an odd function of $\rho$, see figure~\ref{fig:2BodyDecay_RhoDist}. Therefore, the integral of their products vanishes irrespective of the value of $y$.

In a real experiment the velocity of $X$ has a certain distribution $f(\beta)$. Correspondingly, the energy distribution of $l$ in the laboratory frame becomes
\begin{equation}
	D(E_l) ~=~ \int d\beta \,f(\beta) \,\mathcal{D} (E_l\,;\beta) \,,
\end{equation}
where $f(\beta)$ is normalized as $\int d\beta \,f(\beta) = 1$. Even though $D(E_l)$ depends on $f(\beta)$, one finds also in this case
\begin{eqnarray}
	\int dE_l \,D(E_l) \,W(E_l)
		~=~ \int d\beta \,f(\beta) \int dE_l \,\mathcal{D} (E_l\,;\beta) \,W(E_l) ~=~ 0 \,.
	\label{eq:IntegralZero_LabSys}
\end{eqnarray}

\begin{figure}[t]
	\begin{center}
		\begin{tabular}{c}
		
		 	\begin{minipage}{0.5\hsize}
				\begin{center}
					\includegraphics[width=.96\textwidth]{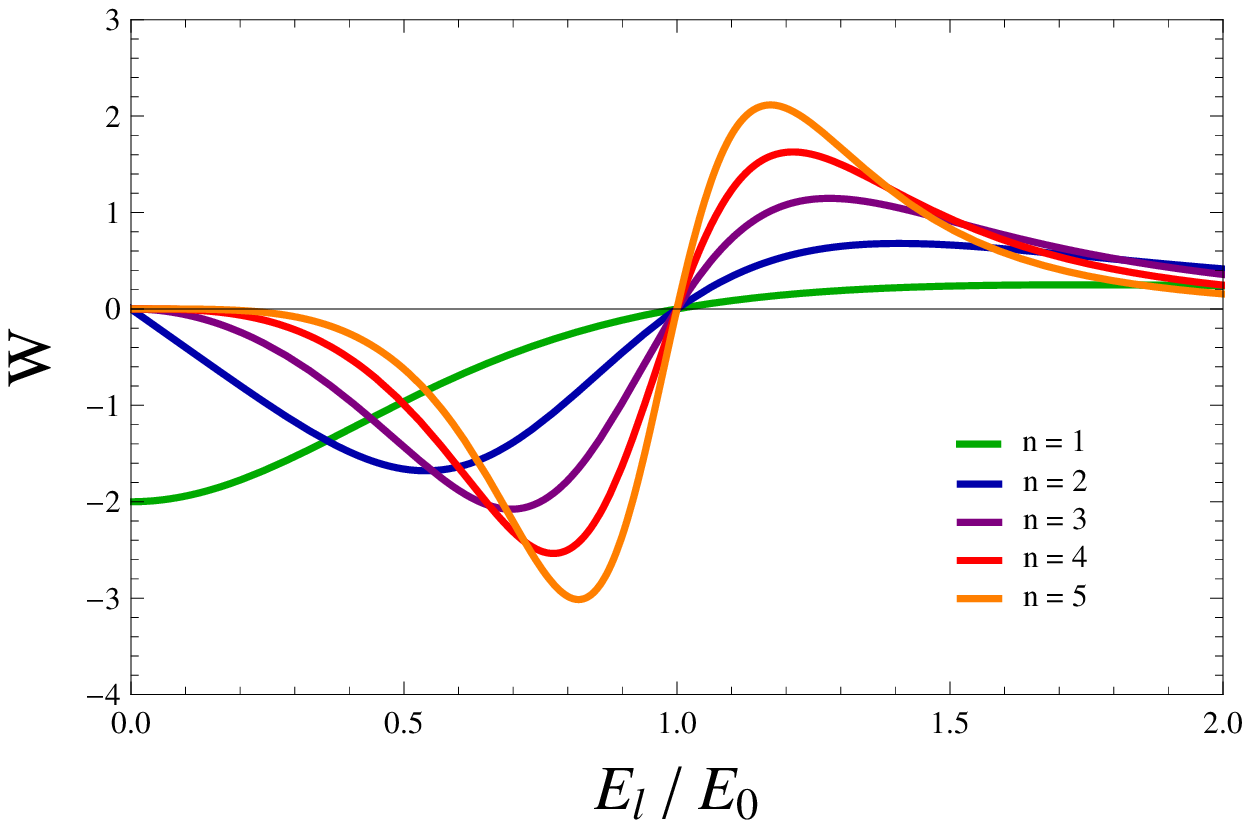}
					\hspace{4.0cm} \small{i) (odd function of $\rho$) $=n\tanh (n\rho)/\!\cosh (n\rho)$}
				\end{center}
			\end{minipage}
			
			\begin{minipage}{0.5\hsize}
				 \begin{center}
					\includegraphics[width=.96\textwidth]{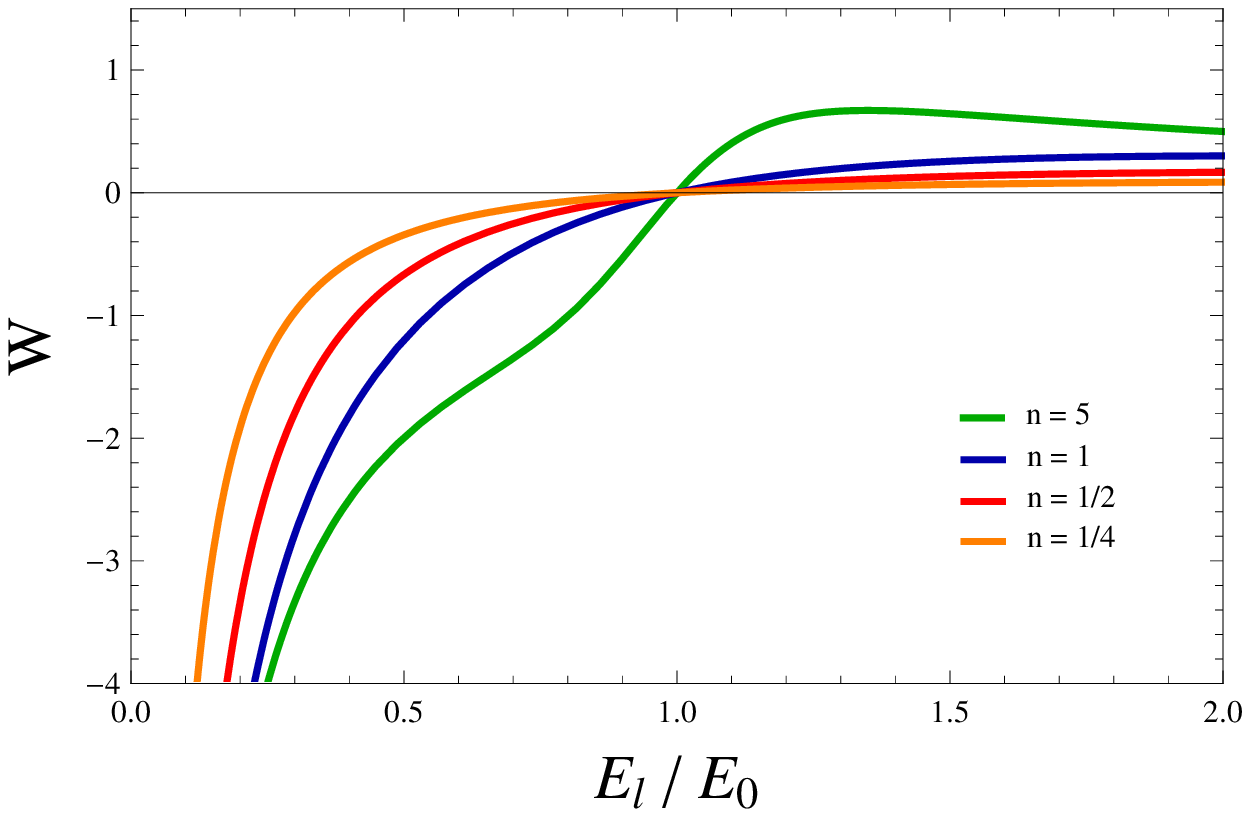}
					\hspace{4.0cm} \small{ii) (odd function of $\rho$) $=\tanh (n\rho)$}
				\end{center}
			\end{minipage}
						
		\end{tabular}	
		\caption{\label{fig:WF_2BodyDecay} Examples of characteristic weight functions for the two-body decay case. $E_0$ and $E_l$ represent the energy of $l$ in the rest frame and laboratory frame of the parent particle, respectively.}		
	\end{center}
\end{figure}

There exist an infinite number of characteristic weight functions, since we can freely choose the odd function of $\rho$ in eq.~\eqref{eq:WF_2BD}. Let us give some examples of characteristic weight functions:
\begin{itemize}

\item[i)] For $(\text{odd function of }\rho)\,=\,n\tanh (n\rho)/\!\cosh(n\rho)$,
\begin{equation}
	W(E_l) ~=~ \left.\frac{2n\,(x^n-x^{-n})}{x\,(x^n+x^{-n})^2}\right|_{x=E_l/E_0} 
		~=~~ \frac{2n\,E_l^{\,n-1}E_0^{\,n+1}\left(E_l^{\,2n}-E_0^{\,2n}\right)}{\left(E_l^{\,2n}+E_0^{\,2n}\right)^2} \,.
\end{equation}
$W(E_l)$ with $n=1,2,3,4,5$ are shown in figure~\ref{fig:WF_2BodyDecay} i). In the cases $n>1$, these weight functions satisfy $W=0$ at $E_l=0$, giving relatively small weight at $E_l \sim 0$ in the integrand of \eqref{eq:IntegralZero}.

\item[ii)] For $(\text{odd function of }\rho)\,=\,\tanh (n\rho)$,
\begin{equation}
	W(E_l) ~=~ \left.\frac{x^n-x^{-n}}{x\,(x^n+x^{-n})}\right|_{x=E_l/E_0} 
		~=~~ \frac{E_0}{E_l}\,\,\frac{\,E_l^{\,2n}-E_0^{\,2n}\,}{E_l^{\,2n}+E_0^{\,2n}} \,.
\end{equation}
$W(E_l)$ with $n=5,1,\frac{1}{2},\frac{1}{4}$ are shown in figure~\ref{fig:WF_2BodyDecay} ii). These weight functions diverge at $E_l=0$, giving large weight at $E_l \sim 0$ when integrated.

\end{itemize}

\subsection{The case of many-body decay}
\label{sec:ManyBodyDecay}

We generalize the argument for the case of two-body decay to the case of many-body decay. Let us consider the decay of a particle $X$ into many bodies including a particle $l$ (\,$X \rightarrow \,l \,+\, \text{anything}$\,). We require the following conditions to this process:
\begin{itemize}
	\item $l$ is massless and its energy distribution is accurately measurable.
	\item $X$ is scalar or unpolarized.
	\item The energy distribution of $l$ in the rest frame of $X$ is known theoretically.
\end{itemize}

To obtain characteristic weight functions for many-body decay, we can use the result of the two-body decay. Using a trivial equation for the normalized energy distribution $\mathcal{D}_0(E_l)$ of $l$ in the rest frame of $X$,
\begin{equation}
	\mathcal{D}_0(E_l) ~=~ \int dE \,\mathcal{D}_0(E) \,\delta(E_l-E) \,,
\end{equation}
one obtains the normalized energy distribution of $l$ in a boosted frame where $X$ has a velocity $\beta$ as
\begin{equation}
	\mathcal{D}(E_l\,;\beta) ~=~ \int dE \,\mathcal{D}_0(E)~ \frac{1}{2E \sinh y} ~\theta \left( e^{-y} E \,\leq\, E_l \,\leq\, e^y E \right) \,.
	\label{eq:LepEneDist_MBD}
\end{equation}

We construct a weight function $W(E_l)$ such that the integral of $\mathcal{D}(E_l\,;\beta)$ weighted by $W(E_l)$ is independent of the parent particle's velocity $\beta$. It is found that such a weight function for many-body decay can be written by that for the two-body decay:
\begin{equation}
	W (E_l) ~=~ \int dE' \mathcal{D}_0(E') \,\frac{1}{E'^{\,2}} \left.\frac{dG(x)}{dx}\right|_{x=E_l/E'} \,,
	\label{eq:obs}
\end{equation}
where $G$ is the same as in the two-body decay: $G(e^y) = (\,\text{even function of $y$}\,) +\text{const.} \times \sinh y$. 
\begin{proof}
Using eqs.~\eqref{eq:LepEneDist_MBD} and \eqref{eq:obs}, the weighted integral becomes
\begin{eqnarray}
	\int dE_l \,\mathcal{D}(E_l) \,W(E_l)
		~=\, \int dE\,dE' \mathcal{D}_0(E) \,\mathcal{D}_0(E')~ \frac{G\left(e^y E/E'\right)-G\left( e^{-y} E/E'\right)}{2\,E\,E'\sinh y} \,.
	\label{eq:Integral_Proof}
\end{eqnarray}
In the case $G(e^y)=\text{(~even function of $y$~)}$, $G$ satisfies $G(x)=G(x^{-1})$, hence
\begin{equation}
	G\left(e^y \frac{E}{E'}\right)-G\left( e^{-y} \frac{E}{E'}\right) ~=~ G\left(e^y\frac{E}{E'}\right)-G\left( e^y \frac{E'}{E}\right)\,
\end{equation}
is antisymmetric under the exchange of $E$ and $E'$. Since the other part of \eqref{eq:Integral_Proof} is symmetric under the exchange of $E$ and $E'$, the weighted integral vanishes.  In the case $G(e^y)=\sinh y$, 
\begin{equation}
	G\left(e^y \frac{E}{E'}\right)-G\left( e^{-y} \frac{E}{E'}\right) 
		~=~ EE'\left( \frac{1}{E^2} +\frac{1}{E'^{\,2}}\right) \,\sinh y \,,
\end{equation}
from which it follows that
\begin{equation}
	\int dE_l \,\mathcal{D}(E_l) \,W(E_l) ~=~ \int dE \,\mathcal{D}_0(E) \,\frac{1}{E^2} ~=~ \text{independent of $y$\,.}
\end{equation}
Thus, in both cases the weighted integral is independent of $\beta$.
\end{proof}
Eq.~\eqref{eq:obs} can be written as
\begin{equation}
	W(E_l) ~=~ \int dE \,\mathcal{D}_0(E) \frac{1}{EE_l}
		\left.\left[ \,\text{(~odd function of $\rho$~)}~+~\text{const.}\times \cosh \rho~\right]\,\right|_{e^{\rho}=E_l/E} \,.
	\label{eq:WF_Odd+Cosh_MBD}
\end{equation}
We can omit the second term in the square bracket in the same way as the two-body decay case.

Finally we obtain the characteristic weight functions for the many-body decay, $X\rightarrow \,l+\text{anything}$, as
\begin{equation}
	W(E_l) ~=~  \int dE \,\mathcal{D}_0(E) \,\frac{1}{E E_l}
		\!\times \!\left.\text{(~odd function of $\rho$~)}\right|_{e^\rho=E_l/E} \,,
	\label{eq:WeightFunc_MBD}
\end{equation}
which satisfy
\begin{equation}
	\int dE_l \,\mathcal{D}(E_l\,;\beta)\,W(E_l) ~=~ 0 \,.
\end{equation}
In the same way as the two-body decay case, when the parent particle has a velocity distribution $f(\beta)$, the energy distribution of $l$ becomes
\begin{equation}
	D(E_l) ~=~ \int d\beta \,f(\beta) \,\mathcal{D} (E_l\,;\beta) \,.
\end{equation}
Even in this case, the weighted integral remains to be zero:
\begin{eqnarray}
	\int dE_l \,D(E_l) \,W(E_l)
		~=~ \int d\beta \,f(\beta) \int dE_l \,\mathcal{D} (E_l\,;\beta) \,W(E_l) ~=~ 0 \,.
	\label{eq:IntegralZero_LabSys_MBD}
\end{eqnarray}

Let us give some examples of the characteristic weight functions:
\begin{itemize}

\item[i)] For $(\text{odd function of }\rho)\,=\,n\tanh (n\rho)/\!\cosh(n\rho)$,
\begin{equation}
	W(E_l) ~=~  \int dE \,\mathcal{D}_0(E) \,\,\frac{2n\,E_l^{n-1} E^{n-1}(E_l^{2n}-E^{2n})}{(E_l^{2n} + E^{2n})^2} \,.
\end{equation}

\item[ii)] For $(\text{odd function of }\rho)\,=\,\tanh (n\rho)$,
\begin{equation}
	W(E_l) ~=~  \int dE \,\mathcal{D}_0(E) \,\frac{E_l^{2n}-E^{2n}}{E_l\,E\,(E_l^{2n} + E^{2n})} \,.
\end{equation}

\end{itemize}

\subsection{Determination of physical parameters using characteristic weight functions}
\label{sec:WFM}

The above characteristic weight functions can be utilized for determination of physical parameters in experiments. Suppose that we intend to measure a parameter $\lambda$ which enters the theoretical formula for $\mathcal{D}_0(E_l)$. We assume that we aim for an accurate measurement of $\lambda$, and hence the decay process and the interaction of the parent particle are already known sufficiently well such that an explicit expression of $\mathcal{D}_0(E_l)$ is inferred. In other words, we have at hand a concrete theoretical model, for which we want to determine $\lambda$ accurately. A typical example of $\lambda$ is the parent particle's mass $m_X$. We use a charged lepton $e$ or $\mu$ for the particle $l$ since their momenta can be measured accurately, although in principle, other particles can also be used. 

Let us define
\begin{eqnarray}
	 I(\lambda) ~\equiv~ \int dE_l \,D(E_l\,;\lambda^{true}) \,W(E_l\,,\lambda) \,,
	 \label{eq:Descri_Chi}
\end{eqnarray}
where $D(E_l\,;\lambda^{true})$ is the lepton energy distribution measured in the laboratory frame, and $W(E_l\,,\lambda)$ is a characteristic weight function defined by \eqref{eq:WeightFunc_MBD}. Here, we take $\lambda$ as a variable, since $W$ has a $\lambda$ dependence through $\mathcal{D}_0(E_l\,;\lambda)$. $\lambda^{true}$ in the arguments of $D$ indicates that the measured distribution knows the true value of $\lambda$. It follows from eq.~\eqref{eq:IntegralZero_LabSys_MBD} that $I(\lambda)$ satisfies
\begin{equation}
	I(\lambda=\lambda^{true}) ~=~ 0 \,.
	\label{eq:DescriZero}
\end{equation}
Therefore, we should look for zeros of $I(\lambda)$ to obtain $\lambda^{true}$.\footnote{Even in the case that there is more than one zero of $I(\lambda)$, the zeros are usually separated sufficiently apart, such that the correct one can be identified using other distributions such as $m_T$ distribution and lepton $p_T$ spectrum. }

In practice, realistic experimental conditions affect the above ideal picture. The lepton energy distribution obtained in an experiment differs from the ideal one due to the limited acceptance of detectors, a series of cuts applied for event selection, and backgrounds which remain after these cuts. Therefore, the zero of $I(\lambda)$ using a realistic distribution is generally different from $\lambda^{true}$. It is necessary that the difference between them should be estimated by Monte Carlo (MC) simulations which take into account the above experimental effects. Since MC prediction have small uncertainties concerning lepton kinematics, we expect the deviation from $\lambda^{true}$ can be estimated accurately. We confirm this feature in a simulation analysis of the Higgs mass reconstruction in section~\ref{sec:HiggsSimulation}. 

Suppose the distribution measured in an experiment $D_{\text{exp}}(E_l\,;\lambda^{true})$ is different from the ideal distribution $D(E_l\,;\lambda^{true})$ by $\delta D(E_l)$:
\begin{equation}
	D_{\text{exp}}(E_l\,;\lambda^{true}) ~=~ D(E_l\,;\lambda^{true}) \,+\, \delta D(E_l) \,.
\end{equation}
In the case $\left| \delta D/ D\right| ~\ll~ 1$, the difference between the zero of $I(\lambda)$ from experiment and $\lambda^{true}$ is given by
\begin{equation}
	\delta \lambda ~=~ -\int dE_l \,\delta D(E_l) \,W(E_l\,,\lambda^{true})\left/
		\,\left(\left.\frac{\partial I}{\partial \lambda}\,\right|_{\lambda=\lambda^{true}}\right.\right) \,,
	\label{eq:DeltaLambda}
\end{equation}
where $D$ and $\delta D$ satisfy $\int dE_l D=\int dE_l (D+\delta D)=1$, and we dropped ${\cal O} ((\delta D)^2)$ corrections. Using this formula, $\delta \lambda$ is obtained from the estimation of $\delta D$ by MC simulations. In the case that the above approximation is not valid, a fit of $I(\lambda)$ using MC prediction is more appropriate. We define
\begin{equation}
	I_{\text{exp}}(\lambda) ~\equiv~ \int dE_l\,D_{\text{exp}}(E_l\,;\lambda^{true})\,W(E_l\,,\lambda) \,,
\end{equation}
and
\begin{equation}
	I_{\text{MC}}(\lambda) ~\equiv~ \int dE_l\,D_{\text{MC}}(E_l\,;\lambda^{\text{MC}})\,W(E_l\,,\lambda) \,,
\end{equation}
where $D_{\text{MC}}(E_l\,;\lambda^{\text{MC}})$ is the lepton energy distribution predicted by MC simulations corresponding to the input $\lambda =\lambda^{\text{MC}}$. We introduce a distance $d$ between $I_{\text{exp}}(\lambda)$ and $I_{\text{MC}}(\lambda)$ as
\begin{equation}
	d^2(\lambda^{\text{MC}}) ~\equiv~ \int_{\lambda_1}^{\lambda_2} d\lambda 
		\left[ \,I_{\text{exp}}(\lambda)-I_{\text{MC}}(\lambda)\,\right]^2 \,.
	\label{eq:Fit_d^2}
\end{equation}
The reconstructed value of $\lambda^{true}$ is given by $\lambda^{\text{MC}}$ which minimizes $d^2$.\footnote{The interval $[\lambda_1,\,\lambda_2]$ is arbitrary. In the case of a reconstruction of the Higgs boson mass, which we will discuss later, the result is only weakly dependent on the choice of this interval except the case that the interval is too small.}

Note that there are an infinite number of characteristic weight functions which we can use. In principle, simultaneous reconstruction of more than one physical parameter is possible using this large degree of freedom, where multi-variable equations \,$I_j\,(\lambda_1, \,\cdots, \lambda_n)=0~(\,j=1,\,\cdots, N\,)$ should be solved. We do not explore this possibility in this paper, however.

\section{Higgs boson mass reconstruction using $H\rightarrow WW \rightarrow l\nu l\nu$ at LHC : Simulation analysis}
\label{sec:HiggsSimulation}

In this section, we examine the reconstruction of the Higgs boson mass at the LHC as an application of the weight function method described in the previous section. We perform a simulation analysis of the $H\rightarrow WW\rightarrow l\nu l\nu \,(l=e, \mu)$ decay mode assuming the true Higgs boson mass to be $m_H=125$\,GeV, and estimate the accuracy of $m_H$ measurement. Through this analysis we reveal characteristic properties of this method.

The Higgs bosons are produced at the LHC via gluon-gluon fusions (ggF) dominantly, and via vector-boson fusions (VBF) subdominantly. In order to maximize the sensitivity of the Higgs boson search, candidate events for the Higgs boson signal in the $WW$ decay mode are categorized according to the number of energetic jets in the final state. In the $H+0$-jet and $H+1$-jet channels, the signal events mostly originate from the ggF process. On the other hand, $H+2$-jet channel contains the signal events mainly from the VBF process. 

At the present stage of the LHC where the data at a center-of-mass energy ($\sqrt{s}$\,) of 8\,TeV have been collected, analyses of the VBF (2-jet) channel are severely limited by few statistics. Thus in this paper, we study the VBF channel with $\sqrt{s}=14$\,TeV (section \ref{sec:VBF}) and the ggF channel with $\sqrt{s}=8$\,TeV (section \ref{sec:ggF}), respectively. In the ggF channel, 1-jet channel is omitted in our study because of complexity of its background analysis and an expected weak sensitivity. 

\subsection{Vector Boson Fusion Channel}
\label{sec:VBF}

\subsubsection{Analysis setup}

The signal and background processes we consider in this analysis are as follows:
\begin{itemize}
	\item $qqH~\rightarrow ~qq\,W^{+}W^{-} ~\rightarrow ~qq\,l^{+}\nu \,l^{-}\overline{\nu}~~~(\,l=e,\mu\,)$
	\item $t\overline{t} ~and~ Wt ~production$
	\item $Electroweak~ WW+jets~ production$.
\end{itemize}
For simplicity, (1) we omit contributions from the ggF process to the signal events which in reality remains partly, even after event selection cuts for the 2-jet channel. (2) Although $W$ boson can decay into electron or muon via tau lepton $W \rightarrow \tau \nu \rightarrow l \nu \nu \,\nu$, we do not include such events. Backgrounds not given in the above list are shown to be relatively small after all the cuts are applied~\cite{Asai:2004ws}.

Both signal and background events are generated using the MadGraph/MadEvents~\cite{Maltoni:2002qb,Alwall:2007st,Alwall:2011uj}  MC event generator with $\sqrt{s}=14$\,TeV, and then passed to PYTHIA~\cite{Sjostrand:2006za} which performs the parton showering and hadronization, including initial-state-radiation (ISR) and final-state-radiation (FSR). We use CTEQ6L~\cite{Pumplin:2002vw} for the parton distribution function (PDF). All generated events are passed to the fast detector simulator PGS~\cite{PGS}. In order to evaluate systematic properties of the weight function method, we generate sufficiently many events such that statistical fluctuations of the MC events can be ignored. 

In our analysis, we generate only events with just two jets in the final state at parton level. The cone algorithm with $R=0.5$ is used for jet reconstruction in PGS. The cross sections are calculated at leading order. For these reasons, the jet multiplicity of the real data and that of the simulated events are expected to be different. In particular, the cross sections would be underestimated.

On these MC events, we impose event selection cuts following those in ref.~\cite{Asai:2004ws}, which investigated the potential for a discovery of the Higgs boson in the VBF process at the ATLAS experiment. Events are categorized into three modes, $ee,~e\mu,~\mu\mu$, corresponding to leptons in the final state. For details of the event selection, see ref.~\cite{Asai:2004ws}. In our analysis, we make three modifications to their cuts:
\begin{enumerate}
	\item Lepton acceptance:\\
		In view of the current status of the LHC experiment, we tighten lepton acceptance cuts as follows:
		\begin{equation*}
			p_{T}^{1}\,>\,25\,\text{GeV}, ~~~~~~~p_{T}^{2}\,>\,10\,\text{GeV},
		\end{equation*}
		where $p_T^1$ and $p_T^2$ are the transverse momenta of the leading and sub-leading leptons, respectively.
	\item Lepton cuts:\\
		Since two leptons produced by $H\rightarrow W W$ tend to be emitted in the same direction, three lepton cuts are imposed in ref.~\cite{Asai:2004ws}: $\Delta \phi_{ll} \leq 1.5, ~\Delta R_{ll} \leq 1.6, ~M_{ll} <65\,$GeV, where $\Delta \phi_{ll}$ is the azimuthal angle between the lepton directions, $\Delta R_{ll}$ is the separation in $\eta-\phi$ plane, and $M_{ll}$ is the invariant mass of the leptons. We replace these cuts by a single Lorentz invariant cut
		\begin{equation*}
			M_{ll}\, < \,50\,\text{GeV}\,.
		\end{equation*}
		By this modification, the signal efficiency changes within a few \%. If we incorporate the same Lorentz invariant cut into the theoretical lepton energy distribution $\mathcal{D}_0(E)$ and use it for the weight functions [see eq.~\eqref{eq:WeightFunc_MBD}], this cut no longer contributes to the deviation of the reconstructed Higgs boson mass from the true value, since the above condition is independent of the velocity of the Higgs boson.
	\item $b$\,-tagging:\\
	For simplicity, we do not simulate $b$\,-tagging and just estimate the efficiencies of $b$\,-vetos as follows:
	\begin{equation*}
		signal : ~1,~~~~~t\overline{t}~and~Wt : ~0.5, ~~~~~WW+jets : ~1\,.
	\end{equation*}
\end{enumerate}
Table~\ref{tab:CrossSectionVBF} summarizes the cross sections for the signal and background events after applying all the cuts.

\begin{table}[tbp]
	\centering
	\begin{tabular}{|l|ccc|}
		\hline 
		&~~~$ee$~~~&~~~$e\mu$~~~&~~~$\mu\mu$~~~\\
		\hline \hline
		signal (fb) & 0.18 & 0.40 & 0.15\\ \hline
		background (fb) & & &\\
		~~~~~$t\overline{t} ~and~ Wt$ & 0.12 & 0.29 & 0.12\\
		~~~~~$WW+jets$& 0.02 & 0.06 & 0.02\\
		\hline
	\end{tabular}
	\caption{\label{tab:CrossSectionVBF} Cross sections after all cuts.}
\end{table}

The weight functions we use in this analysis are
\begin{equation}
	W(E_l\,,m) ~=~  \int dE \,\mathcal{D}_0(E\,;m) \,\,\frac{2n\,E_l^{n-1} E^{n-1}(E_l^{2n}-E^{2n})}{(E_l^{2n} + E^{2n})^2} \,,
	\label{eq:WeightFuncForVBFHiggs}
\end{equation}
for $n=2,\,3,\,4,\,5$, where we have taken $(\text{odd function of }\rho)$ in eq.~\eqref{eq:WeightFunc_MBD} as $n\tanh (n\rho)/\!\cosh(n\rho)$. These weight functions have a property that $W(E_l\,,m)=0$ at $E_l=0$, thus we expect they suppress the effect of lepton $p_T$ cut which deforms significantly the low energy part of the lepton distribution, as we will discuss later.

\begin{figure}[tbp]
\centering 
\includegraphics[width=.45\textwidth]{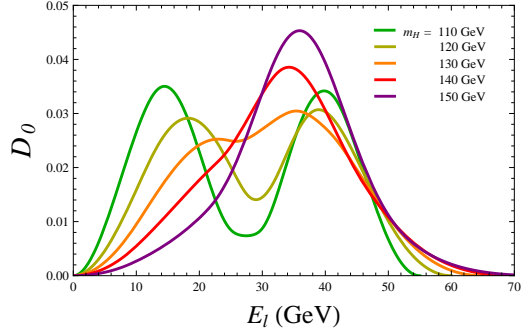}
\caption{\label{fig:TheoreticalLepEneDist} Lepton energy distribution for various Higgs boson masses in the rest frame of the Higgs boson with the restriction $M_{ll}<50$\,GeV. The distribution is normalized as $\int dE_l \,\mathcal{D}_0(E_l\,;m_H)=1$.}
\end{figure}

The normalized lepton energy distribution $\mathcal{D}_0(E_l\,;m)$ in the rest frame of the Higgs boson is given by
\begin{equation}
	\mathcal{D}_0(E_l\,;m_H) ~=~\frac{1}{\Gamma_{M_{ll}<M}}\left.\frac{d\Gamma}{dE_l}\right|_{M_{ll}<M}\,,
\end{equation}
where $\Gamma$ is the decay width for the $H\rightarrow WW \rightarrow l\nu l\nu$ process and a cut on the invariant mass of leptons $M_{ll}<M$ is imposed according to the above discussion ($M=50$ GeV). We obtain
 \begin{eqnarray}
	\left. \frac{d\Gamma}{dE_l}\right|_{M_{ll}<M} &=& \frac{g^6M_W^2}{32\,(2\pi)^5 m_H^2}
				\int_0^{2m_H E_{l}} d\mu^2 \,\frac{1}{(\mu^2-M_W^2)^2+M_W^2\Gamma_W^2}\nonumber \\[0.5ex]
			 &  & ~~\times \,\left[ \,J(-1) ~\theta ( \mu^2 -2 m_H E_{l}+M^2 )
			 	+\,J(z) ~\theta (-\mu^2 + 2m_H E_{l} -M^2 ) \right]\nonumber \\[0.5ex]
			 &  & ~~\times \,\theta (\,m_H -2 E_{l}) \,,
			 \label{eq:LepEneDistWithMll}\\[2ex]
		J(x) &\equiv& \frac{i}{2M_W \Gamma_W} \left[ A_+(x) \log \,({\overline{\mu}}^2_{max}-M_W^2+iM_W \Gamma_W)
					-A_+(x)\log \,(-M_W^2 +iM_W \Gamma_W)\right. \nonumber \\
			&	    & ~~~~~~~~~~~~\left. -A_-(x) \log\,(\overline{\mu}^2_{max} -M_W^2 -iM_W \Gamma_W)
					+A_-(x) \log\, (-M_W^2 -iM_W \Gamma_W) \right] \,,\nonumber \\[1ex]
		A_\pm(x) &\equiv& -\left[ \frac{\mu^2}{8}\left( \frac{1-x^2}{2} +\frac{1-x^3}{3}\right)
					-\frac{1}{16} \left( 2m_H E_l-\mu^2\right) \left( 1-x-\frac{1-x^3}{3} \right)\right]
					(-M_W^2\pm iM_W\Gamma_W)\nonumber \\
			         &          &~~+\frac{1}{16} (2m_H E_l-\mu^2)(m_H^2-2m_H E_l)\left( 1-x-\frac{1-x^3}{3}\right)\,,\nonumber \\[1ex]
		\overline{\mu}^2_{max} &\equiv& \frac{(m_H-2E_l)(2m_HE_l-\mu^2)}{2E_l}\,,~~~
			z ~\equiv~ 1\,-\, \frac{2M^2}{2m_H E_l-\mu^2}\,, \nonumber
\end{eqnarray}
where $g$ is the $SU(2)$ coupling constant, and $M_W$ and $\Gamma_W$ are the mass and total decay width of the $W$ boson, respectively. The parameter $\mu$ represents the invariant mass of an intermediate $W$ boson. Note that $\left.d\Gamma/dE_l\right|_{{\scriptscriptstyle M_{ll}<M}}$ is real. See appendix~\ref{sec:DerivationOfLepEneDist} for the derivation. There remains an integral over $\mu^2$. We carry out this integration numerically and interpolate the numerical table accurately. The obtained theoretical distribution $\mathcal{D}_0(E_l\,;m_H)$ is shown in figure~\ref{fig:TheoreticalLepEneDist} for various Higgs boson masses.

\subsubsection{Mass reconstruction and its sensitivity}
\label{sec:MassReconst}

\begin{figure}[tbp]
	\begin{center}
		\begin{tabular}{c}
		
		 	\begin{minipage}{0.5\hsize}
				\begin{center}
					\includegraphics[width=.92\textwidth]{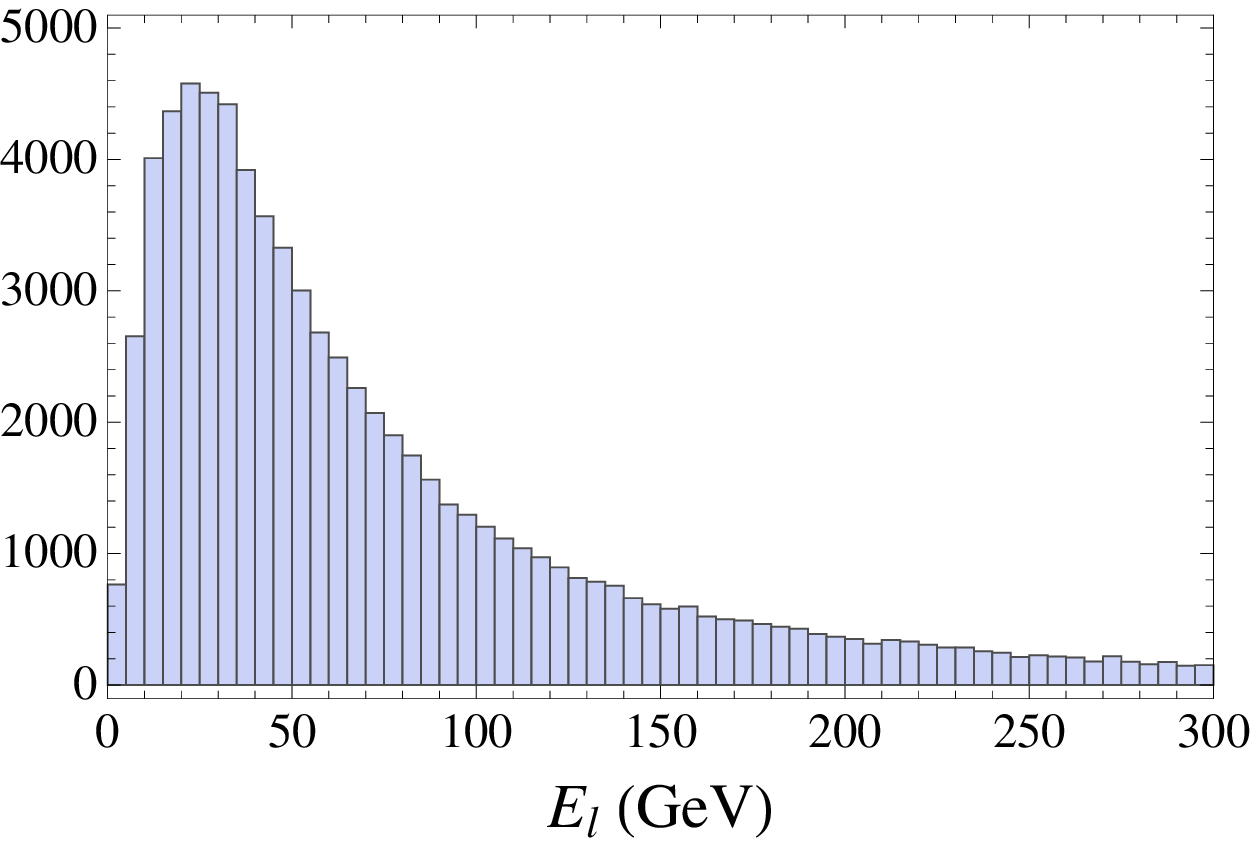}
					\hspace{4.0cm} \small{(a) Parton level (without cuts)}
				\end{center}
			\end{minipage}
			
			\begin{minipage}{0.5\hsize}
				 \begin{center}
					\includegraphics[width=.92\textwidth]{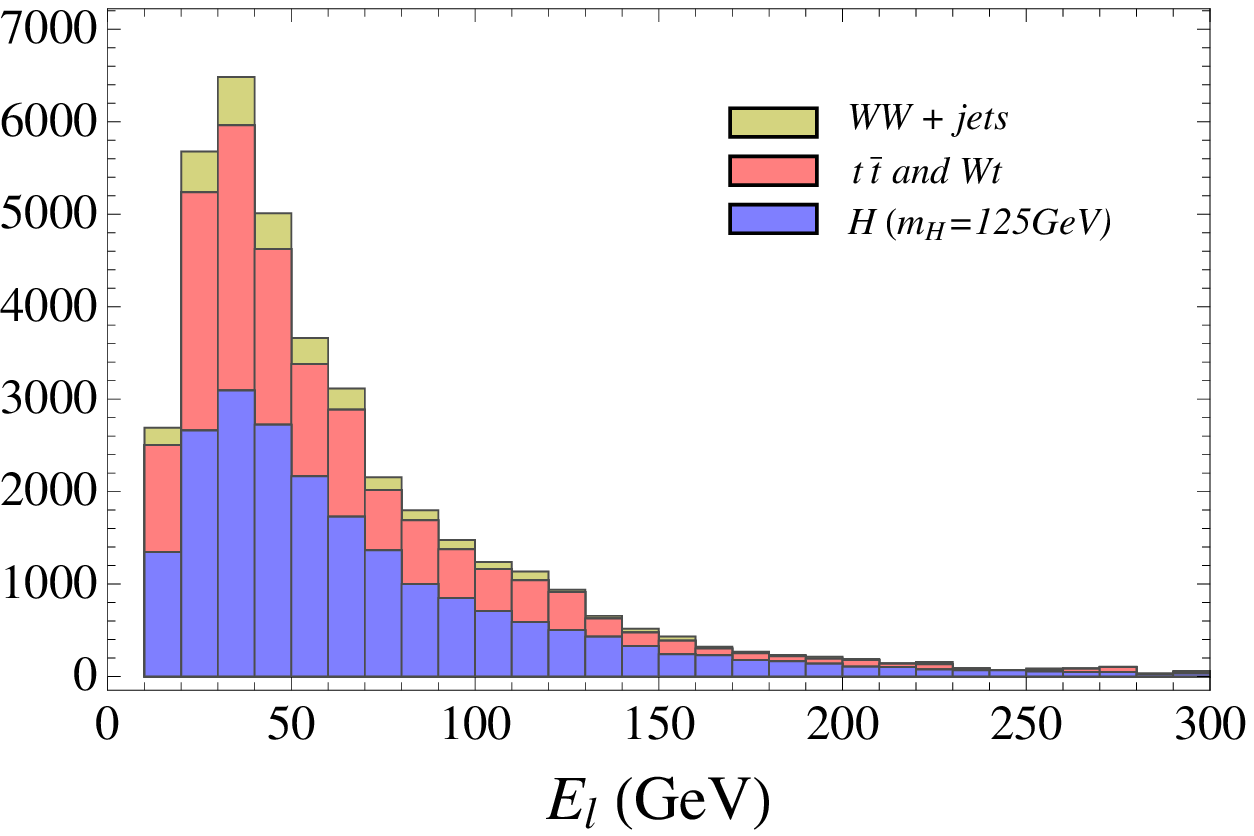}
					\hspace{4.0cm} \small{(b) After all the cuts}
				\end{center}
			\end{minipage}
						
		\end{tabular}	
		\caption{\label{fig:MCLepEneDist} Lepton energy distribution of MC events at parton level (a) and that after all the cuts are applied (b) for a Higgs boson mass of 125\,GeV in the $e\mu$ mode of the VBF channel. Fig.~(a) shows only signal events and (b) includes contributions from background events.}		
	\end{center}
\end{figure}

For reconstruction of the Higgs boson mass, we use the lepton energy distribution of the events which passed all the event selection cuts. Figures~\ref{fig:MCLepEneDist} show the lepton energy distribution of the MC events at parton level (before cuts) (a) and that after all the cuts are applied (b), respectively. Figure~\ref{fig:MCLepEneDist}(a) shows only the signal events, whereas (b) includes contributions from the background events. Comparing figures~\ref{fig:MCLepEneDist}(a) and (b), one can see that the lepton energy distribution after the cuts is deformed especially in the low energy region: $0\leq E_l \lesssim 20$\,GeV. This is mainly due to the lepton $p_T$ cuts in the lepton $p_T$ trigger (\,$p_T^1>25$\,GeV, $p_T^2>10$\,GeV\,).

\begin{figure}[tbp]
\centering 
\includegraphics[width=.45\textwidth]{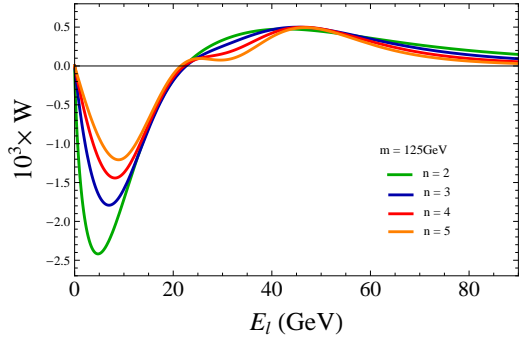}
\caption{\label{fig:WeightFuncForVBFHiggs} Weight functions $W(E_l\,,m)$ at $m=125$\,GeV for the $H\rightarrow WW \rightarrow l\nu l\nu$ decay with the restriction $M_{ll}<50$ GeV.}
\end{figure}

Figure~\ref{fig:WeightFuncForVBFHiggs} shows the weight functions we use in this analysis, defined by eq.~\eqref{eq:WeightFuncForVBFHiggs}. The integration over $E$ in eq.~\eqref{eq:WeightFuncForVBFHiggs} is carried out numerically and we interpolate the numerical table accurately to obtain a valid $W(E_l\,,m)$. In figure~\ref{fig:WeightFuncForVBFHiggs}, one can see that these weight functions give relatively small weight at $E_l\sim 0$, and that this tendency increases with $n$.

\begin{figure}[tbp]
	\begin{center}
		\begin{tabular}{c}
		
		 	\begin{minipage}{0.5\hsize}
				\begin{center}
					\includegraphics[width=.92\textwidth]{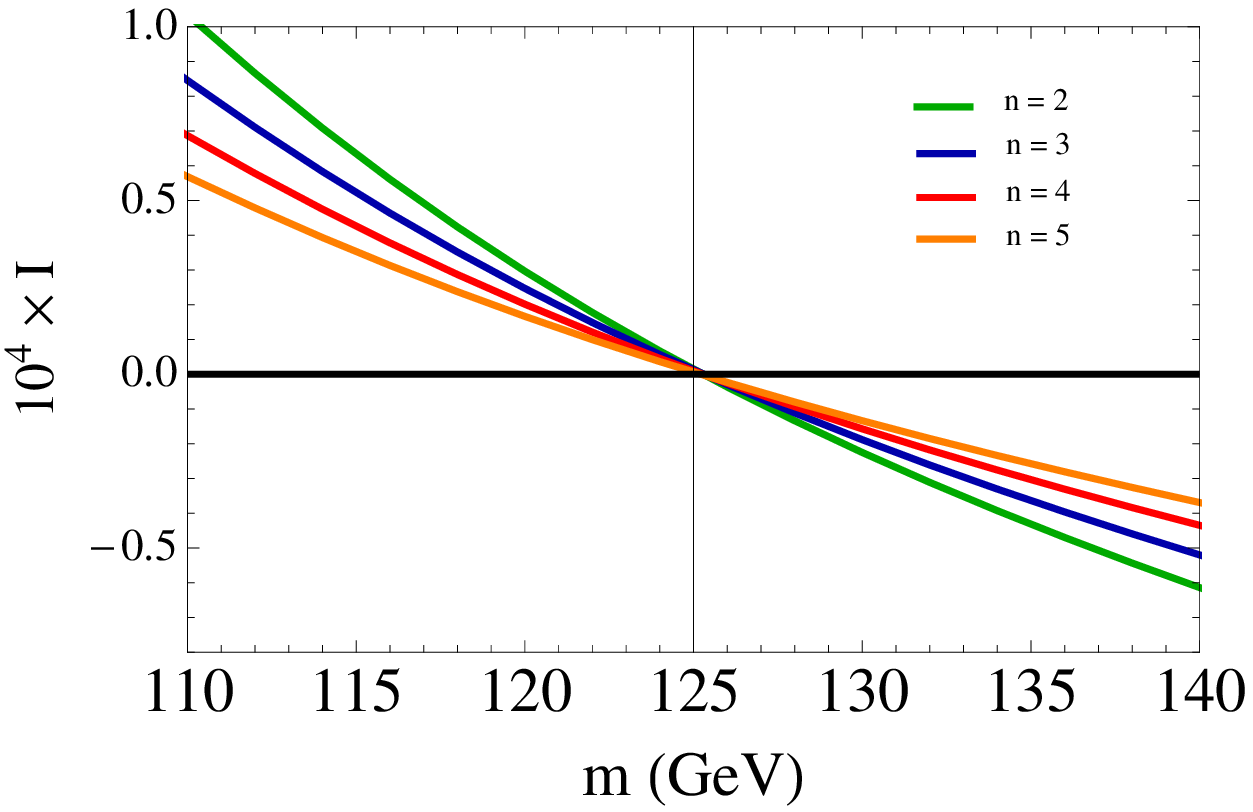}
					\hspace{4.0cm} \small{(a) Parton level (without cuts)}
				\end{center}
			\end{minipage}
			
			\begin{minipage}{0.5\hsize}
				 \begin{center}
					\includegraphics[width=.92\textwidth]{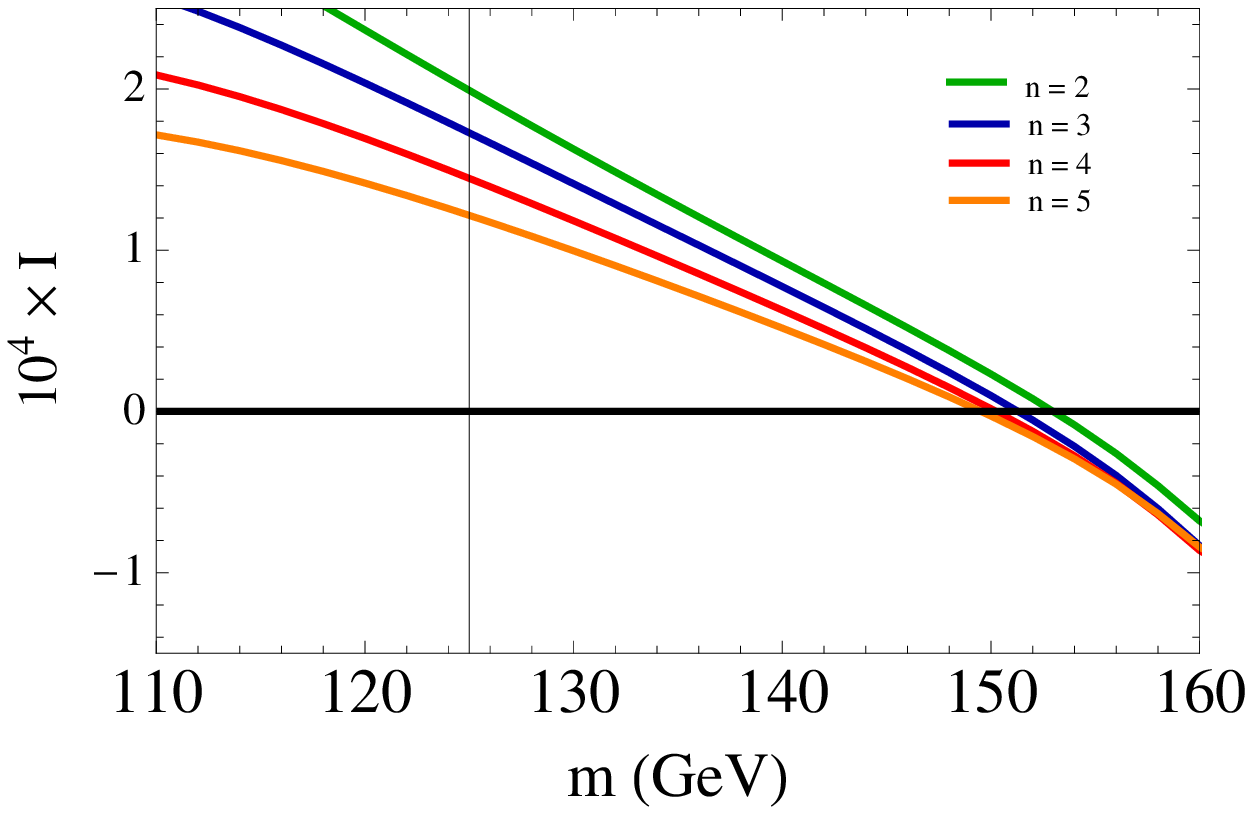}
					\hspace{4.0cm} \small{(b) After all the cuts}
				\end{center}
			\end{minipage}
						
		\end{tabular}	
		\caption{\label{fig:Im_VBF_Cosh} Weighted integral $I(m)$ with lepton energy distribution of MC events at parton level (a) and that after all cuts are applied (b) for an input Higgs mass of 125\,GeV in the $e\mu$ mode. (a) has no contribution from backgrounds, event selection cuts and detector effects, and (b) includes contributions from background events. We use weight functions given in eq.~\eqref{eq:WeightFuncForVBFHiggs} with $n=2,\,3,\,4,\,5$. These lines are computed using the data shown in figures~\ref{fig:MCLepEneDist} and~\ref{fig:WeightFuncForVBFHiggs}.}		
	\end{center}
\end{figure}

With the above lepton energy distribution and weight functions $W(E_l\,,m)$, we construct weighted integrals:
\begin{eqnarray}
	 I(m) ~\equiv~ \int dE_l \,D(E_l\,;m_H) \,W(E_l\,,m) \,,
	 \label{eq:WeightedIntForVBFHiggs}
\end{eqnarray}
where $D(E_l\,;m_H)$ is the normalized lepton energy distribution of the MC events with an input Higgs mass of $m_H$. In figure~\ref{fig:Im_VBF_Cosh}, we show the weighted integrals using the MC lepton distributions at parton level (a) and after all the cuts are applied (b) for $m_H=125$\,GeV. Figure~\ref{fig:Im_VBF_Cosh}(a) confirms $I(m=m_H)=0$, which has been shown in eq.~\eqref{eq:DescriZero}. Therefore, our weight function method works properly in the simulation analysis. By contrast, figure~\ref{fig:Im_VBF_Cosh}(b) indicates that the zero of $I(m)$ is no longer $m_H$ after including the backgrounds and cuts. As we will examine later, the source of the deviation is mainly attributed to the lepton $p_T$ trigger. Since the lepton $p_T$ cut reduces the lepton energy distribution especially in the region $0\leq E_l \lesssim 20$\,GeV, where the weight functions are negative, the weighted integrals $I(m)$ shift in the positive direction from the original one (figure~\ref{fig:Im_VBF_Cosh}(a)). One can see that, using a weight function with larger $n$, this effect by the lepton $p_T$ trigger becomes smaller since the weight function gives a smaller weight at low energy.
Note that the difference between the zero of $I(m)$ and $m_H$ is not sufficiently small compared to $m_H$. For this reason, it is not optimal to reconstruct the mass via eq.~\eqref{eq:DeltaLambda}, which uses the information around the zero of $I(m)$. In this case, following the procedure described in section~\ref{sec:WFM}, we perform a fit of $I(m)$ using MC prediction in order to obtain the reconstructed value.

Let us estimate the sensitivity of the Higgs mass determination. We first estimate the statistical errors, neglecting systematic uncertainties for a moment. Suppose the distribution measured in an experiment $D_\text{exp}(E_l\,;m_H^{true})$ has a statistical fluctuation $\Delta D(E_l)$. Then the reconstructed value of $m_H$ which minimizes $d^2$ defined by \eqref{eq:Fit_d^2} shifts from $m_H^{true}$. This shift is derived from \eqref{eq:Fit_d^2} to be
\begin{equation}
	\Delta m_H ~=~ \int dE_l\, \Delta D(E_l) \,F(E_l)\,,
\end{equation}
with
\begin{eqnarray}
	F(E_l) &\equiv& \frac{1}{A} \,\int dm \,\left[ \,\left.\frac{\partial I_{\text{MC}}\,(m\,;m_H^{\text{MC}})}{ \partial m_H^{\text{MC}}}
		\,\right|_{m_H^{\text{MC}}=m_H^{true}}\,\right] \,W(E_l\,,m)\,,
		\label{eq:F}\\
	A &\equiv& \int dm \,\left[ \,\left.\frac{\partial I_{\text{MC}}\,(m\,;m_H^{\text{MC}})}{ \partial m_H^{\text{MC}}}
		\,\right|_{m_H^{\text{MC}}=m_H^{true}}\,\right]^2\,,
		\label{eq:A}
\end{eqnarray}
where we use the approximation $\left| \Delta m_H/m_H^{true}\right| \ll 1$. Assuming that the statistical fluctuation of the lepton energy distribution follows the Gaussian distribution, one finds the ensemble average of $(\Delta m_H)^2$ to be
\begin{equation}
	\left<(\Delta m_H)^2\right> ~=~ \frac{1}{N} \int dE_l\,D_{\text{MC}}(E_l\,;m_H^{true})
				\left[ \,F(E_l) -\, \int dE_l \,D_{\text{MC}}(E_l\,;m_H^{true})\,F(E_l)\,\right]^2\,,
\end{equation}
where $N$ is the number of leptons. Therefore, we obtain the standard deviation as
\begin{equation}
	\Delta^{stat.} m_H ~=~ \sqrt{\left< (\Delta m_H)^2\right>}
		~~=~~ \frac{~\sqrt{~\overline{\left(F\,-\,\overline{F}\right)^2}~}}{\sqrt{N\,}}\,,
	\label{eq:StatError}
\end{equation}
where we define
\begin{equation}
	\overline{\mathcal{O}} ~\equiv~ \int dE_l \,D_{\text{MC}} (E_l\,;m_H^{true}) \,\mathcal{O}(E_l)\,.
\end{equation}

\begin{table}[tbp]
	\centering
	\begin{tabular}{|cc|ccc|c|}
		\hline 
		\multicolumn{2}{|c}{}&~~~~$ee$~~~~&~~~~$e\mu$~~~~&\multicolumn{2}{c|}{~~~~$\mu\mu$~~~~~~~~~~Combined~~~~}\\
		\hline 
		\multirow{8}{*}{n} & 2 & $+37$ & $+18$ & ~~~$+42$~~~ & $+15~~(\,+8.5\,)$\\ 
		& & $-$ & $-18$ & $-$ & $-18~~(\,-10\,)$\\ 
		\cline{2-6}
		& 3 & $+39$ & $+18$ & $-$ & $+17~~(\,+8.9\,)$\\ 
		& & $-$ & $-18$ & $-$ & $-18~~(\,-10\,)$\\ 
		\cline{2-6}
		& 4 & $+42$ & $+19$ & $-$ & $+18~~(\,+9.4\,)$\\
		& & $-$ & $-19$ & $-$ & $-19~~(\,-10\,)$\\ 
		\cline{2-6}
		&5 & $+44$ & $+20$ & $-$ & $+19~~(\,+9.9\,)$\\ 
		& & $-$ & $-19$ & $-$ & $-19~~(\,-11\,)$\\
		\hline
	\end{tabular}
	\caption{\label{tab:StatError_VBF_1LepTrig} The estimated statistical errors (GeV), $\Delta^{stat.} m_H$, for the Higgs mass reconstruction with an integrated luminosity of 100\,fb$^{-1}$. In brackets, the combined values for 300\,fb$^{-1}$ are also listed. We use the weight functions given in eq.~\eqref{eq:WeightFuncForVBFHiggs} with $n=2,\,3,\,4,\,5$.}
\end{table}

Table~\ref{tab:StatError_VBF_1LepTrig} shows estimates of statistical errors, $\Delta^{stat.} m_H$, for the Higgs mass reconstruction. Some comments are in order: 

(1) To estimate statistical errors, we need to calculate $\left. \partial I_{\text{MC}}/ \partial m_H^{\text{MC}}\right|_{m_H^{\text{MC}}=m_H^{true}}$ [see eqs.~\eqref{eq:F} and \eqref{eq:A}]. We replace the derivative by a finite difference, $[\,I_{\text{MC}}(m\,;m_H^{true}+\delta m_H)- I_{\text{MC}}(m\,;\left. m_H^{true})]\right/ \delta m_H$, with a choice of $\delta m_H$ so as to realize $\Delta^{stat.} m_H \sim \delta m_H$. The integral region of $\lambda = m$ in eqs.~\eqref{eq:Fit_d^2}, \eqref{eq:F} and \eqref{eq:A} is taken as $80$\,GeV$<m<200$\,GeV.\footnote{We find that the statistical error scarcely depends on the choice of the range if we choose a sufficiently wide range (including 125\,GeV), and this range corresponds to one such choice.}

(2) The combined statistical error of all the modes in table~\ref{tab:StatError_VBF_1LepTrig} is defined by
\begin{eqnarray}
	\Delta^{stat.} m_{H,comb.} ~=~ \left[ \frac{1}{\left(\Delta^{stat.}m_{H,\,ee}\right)^2}
	+\frac{1}{\left(\Delta^{stat.} m_{H,\,e\mu}\right)^2}+\frac{1}{\left( \Delta^{stat.}m_{H,\,\mu \mu}\right)^2}\right]^{-\frac{1}{2}}\!\!\!\!,
\end{eqnarray}
where $\Delta^{stat.}m_{H,\,ll}$ for $ll=ee,\,e\mu,\,\mu \mu$ is the statistical error of each $ll$ mode.

(3) `$-$' in table~\ref{tab:StatError_VBF_1LepTrig} shows that it is impossible to find the corresponding value of $\Delta^{stat.}m_H$. This is because the cross section of the signal events in the mass region $m_H \lesssim 110$\,GeV after all the cuts is too small compared with the background events. In this region, the lepton energy distribution is dominated by the backgrounds. Therefore, $I(m\,;m_H^{\text{MC}})$ has little $m_H^{\text{MC}}$ dependence, namely $\partial I_{\text{MC}}/ \partial m_H^{\text{MC}} \sim 0$, which makes it difficult to evaluate $\Delta^{stat.} m_H$. This difficulty should be true also in other methods for the Higgs boson mass reconstruction. We will come back to this point when we discuss the ggF process in section~\ref{sec:ggF}. In addition, since $m_H^{\text{MC}} \sim 160$\,GeV is the turning point of the behavior of $I(m\,;m_H^{\text{MC}})$, reflecting the on-shell $WW$ threshold $2M_W$, $I(m)$ behaves similarly for $m_H^{\text{MC}}$ above 160\,GeV and below 160\,GeV. Hence, the mass region $m_H^{true}+\Delta^{stat.} m_H \gtrsim 165$\,GeV is also a difficult region to evaluate the statistical error. 

(4) The statistical error is smaller for the weight function with a smaller $n$. Qualitatively, this may be understood by noting that the weight function with a smaller $n$ utilizes a wider range of the lepton energy distribution.

Let us now turn to the systematic uncertainties. There are many possible sources, and we examine several major ones, namely uncertainties in the jet energy scale (JES), factorization scale, and $t\overline{t}$ background normalization. Although the $b$\,-tagging efficiency might also be a source of large uncertainties, we do not examine this effect for simplicity.

Suppose that $\mu$ is a parameter with some uncertainty in the MC prediction, and that we choose a value of $\mu$, for example, $\mu^{true}+\Delta \mu$, to reconstruct the Higgs boson mass, where $\mu^{true}$ is the true value of $\mu$ and $\Delta \mu$ represents the order of the uncertainty. Then the reconstructed value of $m_H$ shifts from $m_H^{true}$ by
\begin{equation}
	\Delta^{sys.} m_H ~=~ -\frac{1}{A} \,\int dm \,\frac{\partial I_\text{MC}}{\partial m_H^{\text{MC}}}
	\,\left.\frac{\partial I_{\text{MC}}}{\partial \mu}\right|_{m_H^{\text{MC}}=m_H^{true},\,\mu=\mu^{true}}\times \Delta \mu\,,
	\label{eq:SysError}
\end{equation}
which is derived from eq.~\eqref{eq:Fit_d^2}.

\begin{table}[tbp]
	\centering
	\begin{tabular}{|cc|ccc|}
		\hline 
		\multicolumn{2}{|c}{}&~~~~JES~~~~&~~~Fac. scale~~~&~~$t\overline{t}$ norm.~~\\
		\multicolumn{2}{|c}{}&&(signal)&\\
		\hline 
		\multirow{4}{*}{n} & 2 & $3.1$ & $1.4$ & $0.5$\\ 
		& 3 & $3.5$ & $1.2$ & $0.5$\\ 
		& 4 & $3.8$ & $0.9$ & $0.5$\\
		&5 & $4.0$ & $0.7$ & $0.5$\\ 
		\hline
	\end{tabular}
	\caption{\label{tab:SysError_VBF_1LepTrig} The estimated systematic uncertainties (GeV), $\Delta^{sys.}m_H$, for the Higgs mass reconstruction. We use the weight functions given in eq.~\eqref{eq:WeightFuncForVBFHiggs} with $n=2,\,3,\,4,\,5$, and the MC events in the $e\mu$ mode.}
\end{table}

Table~\ref{tab:SysError_VBF_1LepTrig} shows estimates of systematic uncertainties, $\Delta^{sys.}m_H$, from various sources. In estimating $\Delta^{sys.} m_H$, we replace the derivatives of $I_{\text{MC}}$ in eq.~\eqref{eq:SysError} by finite differences, in the same way as in the estimation of the statistical errors. The integral region of $m$ is taken to be 80\,GeV$\,<m<\,$200\,GeV. We estimate the uncertainties associated with JES by varying the $p_T$ of all the jets in the events by $\pm 10$\,\% before the cuts are applied. Note that we vary the missing $p_T$ simultaneously to conserve the total transverse momentum. The uncertainties from factorization scale is estimated by varying the scale in PDF and PYTHIA by $1/2$ and 2 only for the signal events. In addition, effects due to the uncertainties of $t\overline{t}$ background normalization are evaluated by changing the normalization by $\pm 10$\,\%. It is found that the uncertainty attributed to JES is relatively large compared to other sources. This results mainly from the JES uncertainty of the $t\overline{t}$ background.

Comparing tables~\ref{tab:StatError_VBF_1LepTrig} and \ref{tab:SysError_VBF_1LepTrig}, it is evident that the statistical errors dominate over the systematic errors. Thus, the accuracy of the Higgs boson mass determination in this analysis is limited by statistics.

\subsubsection{Properties of the weight function method}

In this section, we examine the analysis presented in the previous section in more detail to reveal characteristic properties of the weight function method. We investigate the $m_H$ dependence and the effect of the lepton $p_T$ trigger, both of which severely affect the sensitivity of the Higgs mass determination. Also, we comment on the effect of backgrounds.

The branching ratio of the $H \rightarrow WW$ decay mode is a crucial factor in discussing the $m_H$ dependence of the determination of $m_H$. Since this ratio falls off sharply as the value of $m_H$ decreases below the $WW$ threshold $2M_W$, the number of events we can use for the mass reconstruction also reduces for small $m_H$. In the case of $m_H=130$\,GeV, for example, the signal cross section after all the cuts is 1.5 times as large as  that for $m_H=125$\,GeV, which results in the signal-to-background ratio improving by a factor 1.3. In addition, the efficiency of the lepton $p_T$ cut for the signal events also improves in the case for $m_H=130$\,GeV, since leptons tend to be more energetic than those in the case of $m_H=125$\,GeV.

Table~\ref{tab:StatError_VBF_mH130} lists estimates of statistical errors, $\Delta^{stat.} m_H$, for the Higgs boson mass determination with an input mass value of $m_H=130$\,GeV. We find that the statistical errors reduce in all the lepton mode as compared to the case of $m_H=125$\,GeV (see table~\ref{tab:StatError_VBF_1LepTrig}). The relative accuracy of the Higgs mass determination with 100\,fb$^{-1}$ data is estimated to be $+12\%$, $-14\%$ for $m_H=125$\,GeV and improves to $\pm10\%$ for $m_H=130$\,GeV. Moreover, it reaches $\pm2\%$ for $m_H=150$\,GeV as shown in our previous letter~\cite{Kawabata:2011gz}. 
We consider that this strong dependence on $m_H$ results from the growth of the branching ratio and the improvement of the efficiency of the lepton $p_T$ cut as $m_H$ increases. We mention that the possibility of $m_H \sim 130$\,GeV may still remain in view of the recent experimental data~\cite{CMS:2013Morio, ATLAS:2013Comb}.

\begin{table}[tbp]
	\centering
	\begin{tabular}{|ccc|c|}
		\hline 
		~~~~$ee$~~~~&~~~~$e\mu$~~~~&\multicolumn{2}{c|}{~~~~$\mu\mu$~~~~~~~~~~Combined~~~~}\\
		\hline 
		$+29$ & $+17$ & ~~~$+32$~~~ & $+13~~(\,+7.0\,)$\\ 
		$-$ & $-13$ & $-$ & $-13~~(\,-6.4\,)$\\ 
		\hline
	\end{tabular}
	\caption{\label{tab:StatError_VBF_mH130} The estimated statistical errors (GeV), $\Delta^{stat.} m_H$, for the Higgs mass reconstruction at an integrated luminosity of 100\,fb$^{-1}$ with an input Higgs boson mass value of $m_H=$130\,GeV for comparison. In brackets, the combined values for 300\,fb$^{-1}$ are also listed. We use the weight function given in eq.~\eqref{eq:WeightFuncForVBFHiggs} with $n=2$.}
\end{table}

\begin{figure}[tbp]
	\begin{center}
		\begin{tabular}{c}
		
		 	\begin{minipage}{0.5\hsize}
				\begin{center}
					\includegraphics[width=.92\textwidth]{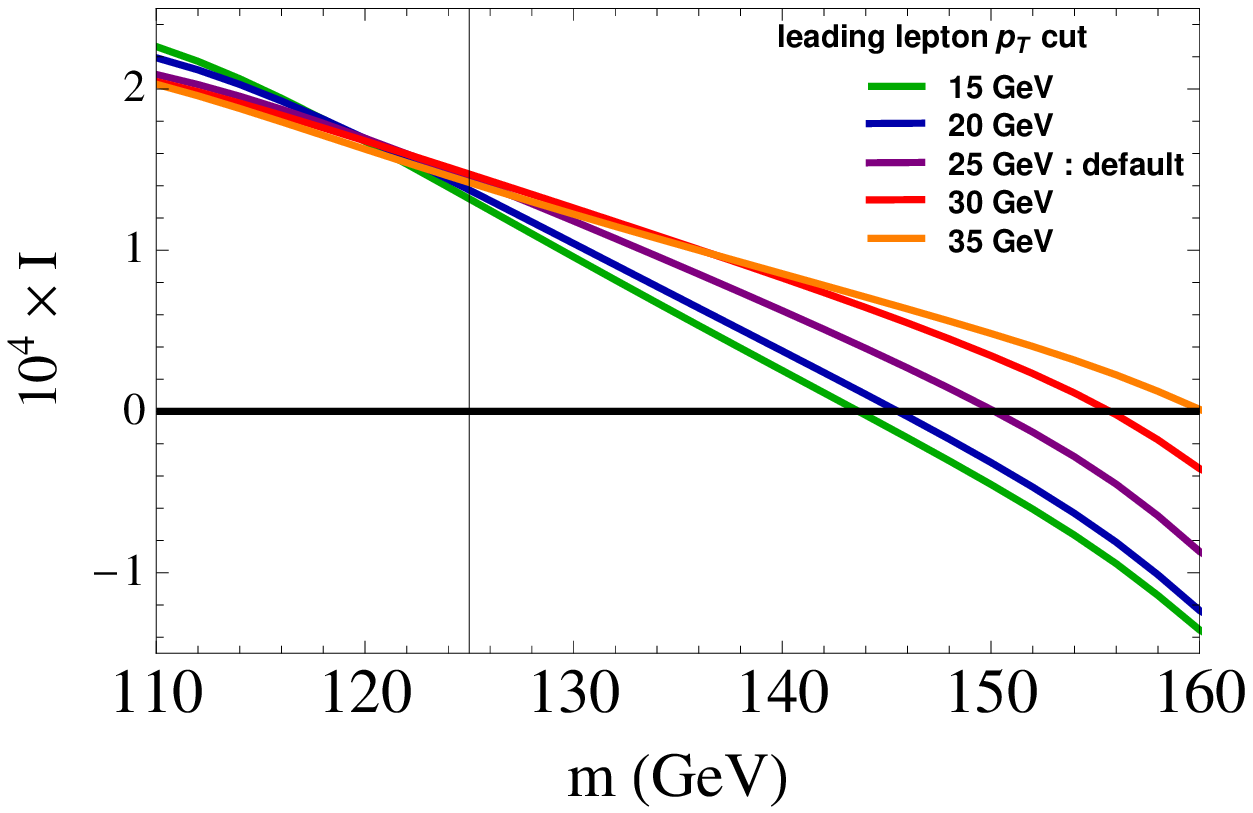}
					\hspace{4.0cm} \small{(a) Lepton $p_T$ cut}
				\end{center}
			\end{minipage}
			
			\begin{minipage}{0.5\hsize}
				 \begin{center}
					\includegraphics[width=.92\textwidth]{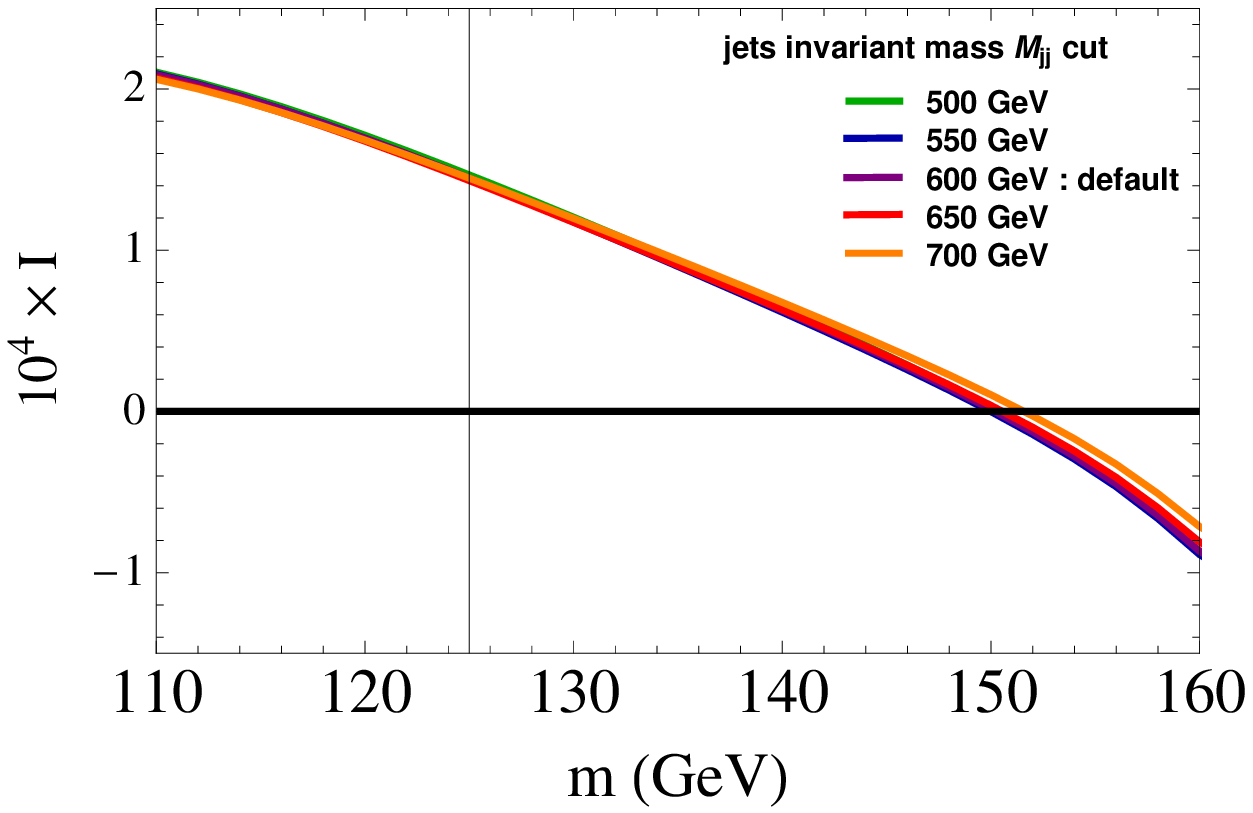}
					\hspace{4.0cm} \small{(b) Jet cut}
				\end{center}
			\end{minipage}
						
		\end{tabular}	
		\caption{\label{fig:Im_VaryCuts} Weighted integrals $I(m)$ with the lepton $p_T$ cut varied (a) and those with the jet cut varied (b) for an input Higgs mass of 125\,GeV in the $e\mu$ mode. 
		We use the weight function given in eq.~\eqref{eq:WeightFuncForVBFHiggs} with $n=4$.}		
	\end{center}
\end{figure}

We now discuss the effect of the lepton $p_T$ trigger. Figure~\ref{fig:Im_VaryCuts}(a) shows the weighted integrals $I(m)$, where the value of the lepton $p_T$ cut for the leading leptons is varied in steps of 5\,GeV. All the other cuts are applied with the same values as before, and contributions from the background events are included. One can see in figure~\ref{fig:Im_VaryCuts}(a) that $I(m)$ depends strongly on the value of the lepton $p_T$ cut. For comparison, we show in figure~\ref{fig:Im_VaryCuts}(b) $I(m)$ as the value of the cuts concerning jets is varied. We vary the lower limit of the invariant mass $M_{jj}$ of the two tagged jets in steps of 50\,GeV. In contrast to the case for varying the lepton $p_T$ cut, $I(m)$ is very stable against the changes of the cut value concerning jets. This is because the jet cut changes only the normalization of the lepton energy distribution and scarcely changes its shape, whereas the lepton $p_T$ cut directly deforms the lepton energy distribution especially in the low energy region.

At the LHC experiments, there is another choice for the lepton trigger, namely, the di-lepton trigger. Here, we examine
\begin{eqnarray*}
	\begin{aligned}
		\begin{cases}
			&p_T(e)\,>\,15\, \text{GeV}  ~~~~~~~~~~~~~~~~~~~~~~~~~~~~~~~~~~~~~\text{for $ee$ mode}\\
			&p_T(e)\,>\,10\, \text{GeV},~ p_T(\mu)\,>\,6\,\text{GeV}~~~~~~~~~~~~~~ \text{for $e\mu$ mode}\\
			&p_T(\mu)\,>\,10\, \text{GeV} ~~~~~~~~~~~~~~~~~~~~~~~~~~~~~~~~~~~~\,\text{for $\mu\mu$ mode\,,}
		\end{cases}
	\end{aligned}
\end{eqnarray*}
where $p_T(l)$ is the transverse momentum of lepton $l$\,\,($=e,\, \mu$). The threshold values of $p_T(l)$ are chosen according to \cite{Masubuchi:PriCom}. Although di-lepton triggers are not used at present in the LHC experiments, it is likely that this type of trigger will be used in future LHC runs~\cite{Masubuchi:PriCom}. The cuts in the di-lepton trigger are looser (especially in the $e\mu$ mode) than the single-lepton trigger which we have used in the analysis. Therefore, we expect that the sensitivity of the Higgs boson mass determination will be improved by using the di-lepton trigger. Despite the fact that the signal-to-background ratio for the di-lepton trigger is worse than that for the single-lepton trigger, we show in table~\ref{tab:StatError_VBF_2LepTrig} that the statistical errors are indeed reduced when using the di-lepton trigger. This fact indicates that the sensitivity of the lepton distribution to $m_H$ is stronger at low energy region where the lepton triggers mainly affect. In view of this feature, we stress importance of using the di-lepton trigger in order to obtain a better accuracy.

\begin{table}[tbp]
	\centering
	\begin{tabular}{|ccc|c|}
		\hline 
		~~~~$ee$~~~~&~~~~$e\mu$~~~~&\multicolumn{2}{c|}{~~~~$\mu\mu$~~~~~~~~~~Combined~~~~}\\
		\hline 
		 $-$ & $+16$ & ~~~$+36$~~~ & $+14~~(\,+7.7\,)$\\ 
		 $-$ & $-17$ & $-$ & $-17~~(\,-7.8\,)$\\ 
		\hline
	\end{tabular}
	\caption{\label{tab:StatError_VBF_2LepTrig} The estimated statistical errors (GeV), $\Delta^{stat.} m_H$, for the Higgs mass reconstruction using di-lepton trigger at an integrated luminosity of 100\,fb$^{-1}$. In brackets, the combined values for 300\,fb$^{-1}$ are also listed. We use the weight function given in eq.~\eqref{eq:WeightFuncForVBFHiggs} with $n=2$.}
\end{table}

We also mention effects of backgrounds. Contributions from the backgrounds decreases the relative significance of the signal events, which results in a worse sensitivity of the lepton distribution to $m_H$. A typical example where this effect becomes crucial is in the ggF process, which we discuss in the next section. Since the signal-to-background ratio gets worse with decreasing $m_H$, a small $m_H$ is disadvantageous also from this viewpoint. The detailed discussion on backgrounds effects is given in the analysis of the ggF channel (section~\ref{sec:ggF}).

\subsection{Gluon Fusion Channel}
\label{sec:ggF}

The analysis setup for the ggF process basically follows that for the VBF process explained in section~\ref{sec:VBF}. Thus, we list only the differences from the VBF analysis. 

The signal and background processes we consider are as follows:
\begin{itemize}
	\item $gg~\rightarrow ~H~\rightarrow ~W^{+}W^{-} ~\rightarrow ~l^{+}\nu \,l^{-}\overline{\nu}~~~(\,l=e,\mu\,)$~~~~:~~$H+0$-jet channel\,
	\item $pp~\rightarrow~WW~\,production$
\end{itemize}
We set $\sqrt{s}=8$\,TeV and generate only the $WW$ background events with no jets in the final state at parton level. For the event selection, successive cuts summerized in table~\ref{tab:ggFCuts} are applied, following refs.~\cite{ATLAS:2011kha} and~\cite{Masubuchi:PriCom}. See ref.~\cite{ATLAS:2011kha} for the detailed definition of the cuts listed in this table. The cross sections for the signal and background events after all the above cuts are summarized in table~\ref{tab:CrossSectionggF}. Note that since these cross sections are calculated at leading order, they are likely to be underestimated especially for the ggF signal.

\begin{table}[tbp]
	\centering
	\begin{tabular}{|l|c|}
		\hline 
		1. Lepton acceptance & ~~Exactly two isolated, opposite-sign leptons ($ee,\,e\mu,\,\mu \mu$) with~~~~\\
		& $p_T^{leading}(l)>25$\,GeV, ~$p_T^{subleading}(l)>10$\,GeV, ~$|\eta(l)|<2.5$\\ \hline
		2. Dilepton invariant mass& $10$\,GeV$<M_{ll}<50$\,GeV ~for $e\mu$ mode\\ 
		& $12$\,GeV$<M_{ll}<50$\,GeV ~for $ee$ and $\mu \mu$ modes\\ \hline
		3. $E_{T,rel}^{miss}$ cut& $E_{T,rel}^{miss}>30$\,GeV\\ 
		4. Jet veto& No jets with $p_T>25$\,GeV in $|\eta|<4.5$\\ 
		5. Dilepton $\bm{p}_T$ cut& $|\bm{p}_T^{ll}|>30$\,GeV\\ 
		6. $\Delta \phi_{ll}$ cut & $|\Delta \phi_{ll}|<1.3$\\ 
		7. Transverse mass & 0.75\,\,$m_H<M_T<m_H$\\ 
		\hline
	\end{tabular}
	\caption{\label{tab:ggFCuts} Successive cuts applied in the ggF (0-jet channel) analysis.}
\end{table}

\begin{table}[tbp]
	\centering
	\begin{tabular}{|l|ccc|}
		\hline 
		&~~~$ee$~~~&~~~$e\mu$~~~&~~~$\mu\mu$~~~\\
		\hline 
		signal (fb) & 1.1 & 2.1 & 0.9\\ 
		background : $WW$ (fb) & 8.8 & 16.2 & 7.1\\
		\hline
	\end{tabular}
	\caption{\label{tab:CrossSectionggF} Cross sections after all the cuts.}
\end{table}

We use the same weight functions as in the VBF analysis~\eqref{eq:WeightFuncForVBFHiggs} except for the constraints on the dilepton invariant mass $M_{ll}$. Corresponding to the value of $M_{ll}$ cut in the ggF analysis, we use the theoretical lepton energy distribution $\mathcal{D}_0(E\,;m)$ in eq.~\eqref{eq:WeightFuncForVBFHiggs} with the restriction $M_1<M_{ll}<M_2$, where $M_1=10\,(12)$\,GeV for $e\mu$\,($ee,\, \mu \mu$) mode and $M_2=50$\,GeV:
\begin{equation}
	\mathcal{D}_0(E_l\,;m_H) ~=~\frac{1}{\Gamma_{M_1<M_{ll}<M_2}}\left.\frac{d\Gamma}{dE_l}\right|_{M_1<M_{ll}<M_2}\,.
\end{equation}
After calculations similar to the VBF analysis, we obtain
 \begin{eqnarray}
	\left. \frac{d\Gamma}{dE_l}\right|_{M_1<M_{ll}<M_2} &=& \frac{g^6M_W^2}{32\,(2\pi)^5 m_H^2} 
				\int_0^{2m_H E_{l}} d\mu^2 \,\frac{1}{(\mu^2-M_W^2)^2+M_W^2\Gamma_W^2}\nonumber \\[0.5ex]
			 &  & ~\times \,\left[ \,\tilde{J}(-1) ~\theta ( M_1^2 < 2 m_H E_{l} - \mu^2 < M_2^2 )
			 	+\,\tilde{J}(z_2) ~\theta (-\mu^2 + 2m_H E_{l} -M_2^2 ) \right]\nonumber \\[0.7ex]
			 &  & ~\times \,\theta (\,m_H -2 E_{l})\,,
			 \label{eq:LepEneDistWithMll}
\end{eqnarray}
\vspace{-0.9cm}
\begin{eqnarray}
		\tilde{J}(x) &\equiv& \frac{i}{2M_W \Gamma_W} 
					\left[ \tilde{A}_+(x) \log \,({\overline{\mu}}^2_{max}-M_W^2+iM_W \Gamma_W)
					-\tilde{A}_+(x)\log \,(-M_W^2 +iM_W \Gamma_W)\right. \nonumber \\
			&	    & ~~~~~~~~~~~~\left. -\tilde{A}_-(x) \log\,(\overline{\mu}^2_{max} -M_W^2 -iM_W \Gamma_W)
					+\tilde{A}_-(x) \log\, (-M_W^2 -iM_W \Gamma_W) \right]\,,\nonumber \\[1ex]
		\tilde{A}_\pm(x) &\equiv& -\left[ \frac{\mu^2}{8}\left( \frac{z_1^2-x^2}{2} +\frac{z_1^3-x^3}{3}\right)
					-\frac{1}{16} \left( 2m_H E_l-\mu^2\right) \left( z_1-x-\frac{z_1^3-x^3}{3} \right)\right]
					(-M_W^2\pm iM_W\Gamma_W)\nonumber \\
			         &          &~~+\frac{1}{16} (2m_H E_l-\mu^2)(m_H^2-2m_H E_l)\left( z_1-x-\frac{z_1^3-x^3}{3}\right)\,,\nonumber \\[1ex]
		z_1 &\equiv& 1\,-\, \frac{2M_1^2}{2m_H E_l-\mu^2}\,,~~~
			 z_2 ~\equiv~ 1\,-\, \frac{2M_2^2}{2m_H E_l-\mu^2}\,.\nonumber
\end{eqnarray}
Using this expression, we can construct weight functions by eq.~\eqref{eq:WeightFuncForVBFHiggs} for reconstructing the Higgs boson mass.

\begin{figure}[tbp]
\centering 
\includegraphics[width=.45\textwidth]{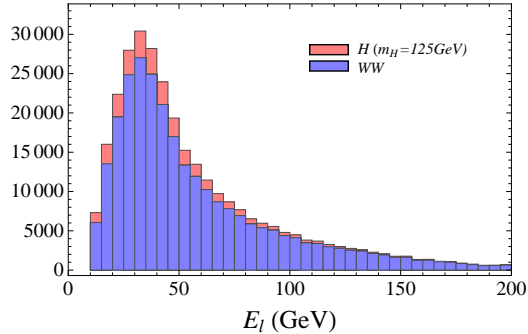}
\caption{\label{fig:MCLepEneDist_ggF} Lepton energy distribution of MC events after all cuts are applied for a Higgs boson mass of 125\,GeV in the $ee$ mode of the ggF channel. The signal and $WW$ background events are piled up.}
\end{figure}

Figure~\ref{fig:MCLepEneDist_ggF} shows the lepton energy distribution of the MC events after all the cuts are applied. The lepton distribution is overwhelmed by the background events. Due to the characteristic feature of the ggF process (0-jet) that there is only the Higgs boson in the final state, this process is contaminated by large backgrounds, unlike the VBF process. 

\begin{figure}[tbp]
\centering 
\includegraphics[width=.45\textwidth]{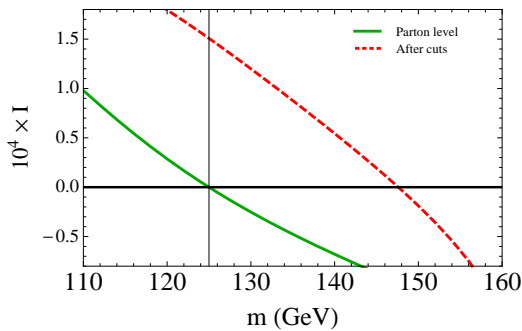}
\caption{\label{fig:Im_ggF_mH125EE} Weighted integral $I(m)$ with MC lepton energy distribution at parton level (the solid line) and that after all cuts are applied (the dashed line) for a Higgs boson mass of 125\,GeV in the $ee$ mode. We use the weight function given in eq.~\eqref{eq:WeightFuncForVBFHiggs} with $n=4$.}
\end{figure}

We construct the weighted integrals $I(m)$ defined by eq.~\eqref{eq:WeightedIntForVBFHiggs} with the weight functions $W(E_l\,,m)$ and the MC lepton energy distribution. Figure~\ref{fig:Im_ggF_mH125EE} shows $I(m)$ at parton level and that after all the cuts are applied. $I(m)$ at parton level has only the signal contribution, which is thus an ideal one, whereas $I(m)$ after all the cuts includes both signal and background contributions. The zero of $I(m)$ at parton level indicates that the input value of the Higgs boson mass $m_H=125$\,GeV is reconstructed correctly using the ideal lepton distribution. On the other hand, the zero of $I(m)$ after the cuts is shifted from $m_H$ due to the effects of the backgrounds and cuts, as in the VBF analysis.

\begin{figure}[tbp]
	\begin{center}
		\begin{tabular}{c}
		
		 	\begin{minipage}{0.5\hsize}
				\begin{center}
					\includegraphics[width=.92\textwidth]{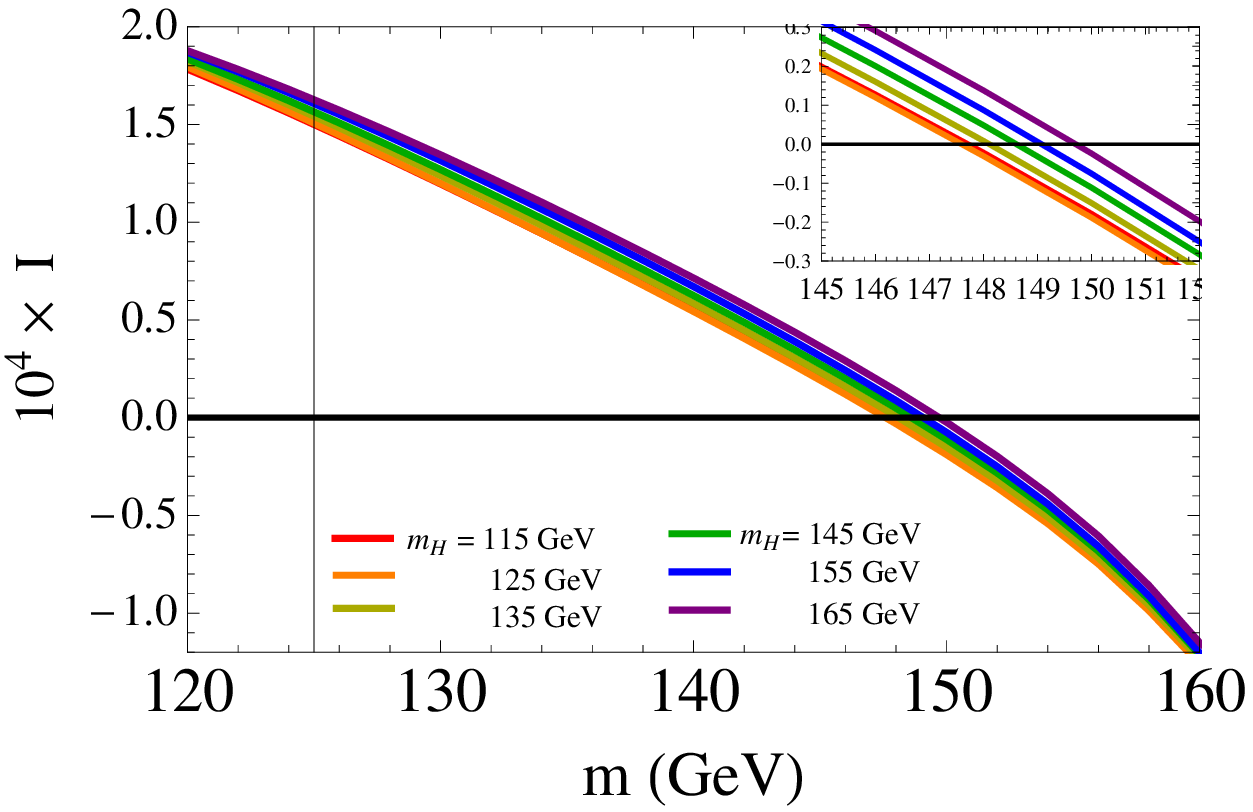}
					\hspace{4.0cm} \small{(a) $I(m)$}
				\end{center}
			\end{minipage}
			
			\begin{minipage}{0.5\hsize}
				 \begin{center}
					\includegraphics[width=.92\textwidth]{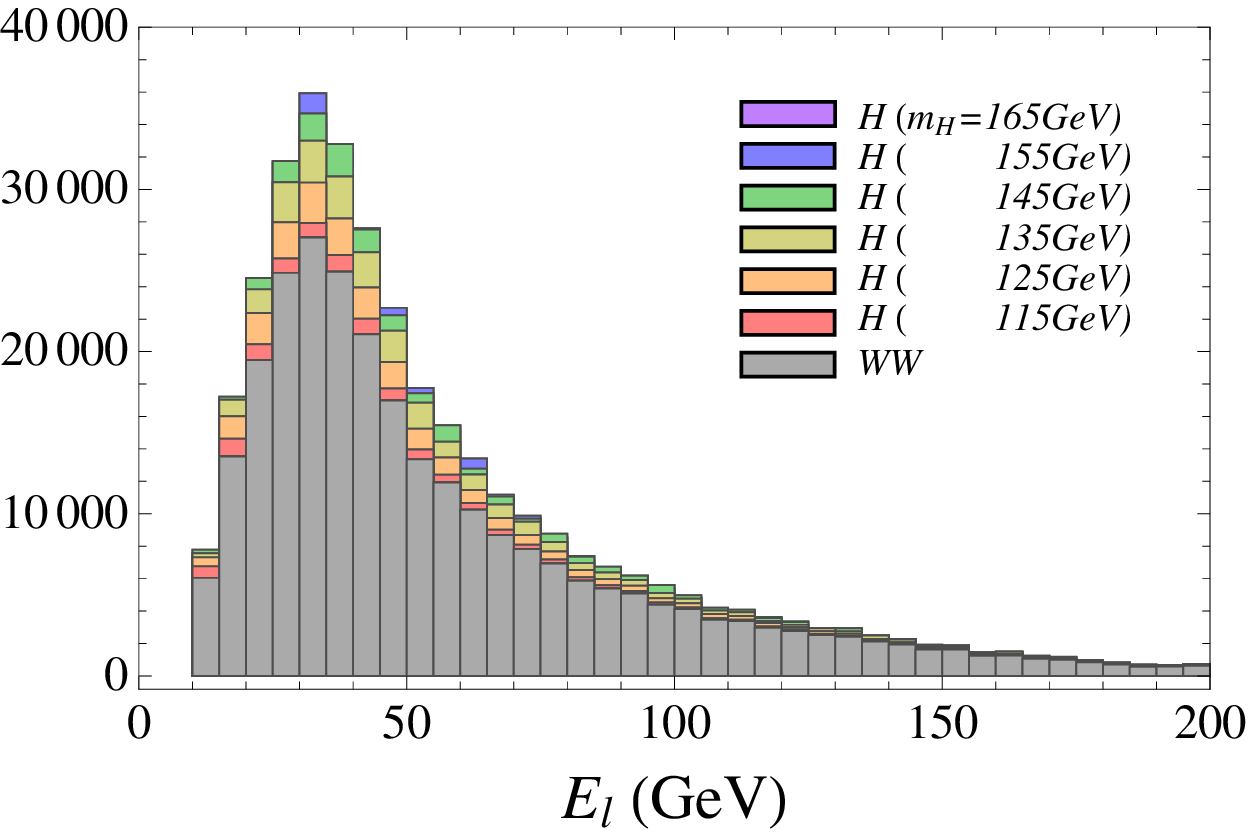}
					\hspace{4.0cm} \small{(b) Lepton energy distribution}
				\end{center}
			\end{minipage}
						
		\end{tabular}	
		\caption{\label{fig:Im_ggF_Cosh} Weighted integral $I(m)$ (a) and lepton energy distribution (b) with MC events after cuts for various input Higgs masses in the $ee$ mode. For (a), we use the weight function given in eq.~\eqref{eq:WeightFuncForVBFHiggs} with $n=4$. For (b), the signal and $WW$ background events are piled up.}
	\end{center}
\end{figure}

Let us investigate the $m_H^{\text{MC}}$ dependence of $I(m)$ which is related to the sensitivity of mass reconstruction when a fitting of $I(m)$ is performed. $I(m)$ corresponding to various input $m_H^{\text{MC}}$ are shown in figure~\ref{fig:Im_ggF_Cosh}(a). One can see that the lines for $I(m)$ are so close to each other especially for small Higgs boson masses that one cannot distinguish $m_H^{\text{MC}}=115$\,GeV and 125\,GeV even in an enlarged view. In order to see the reason for this closeness, we show in figure~\ref{fig:Im_ggF_Cosh}(b) the lepton energy distribution including $WW$ background after all the cuts are applied, for various $m_H^{\text{MC}}$. The shape of the lepton energy distribution is almost determined by the background. Consequently, it is only weakly sensitive to the value of $m_H^{\text{MC}}$, especially for small $m_H^{\text{MC}}$. For $m_H^{\text{MC}}<115$\,GeV, $I(m)$ scarcely changes because the signal events give almost no contribution to the lepton energy distribution compared to backgrounds. This means that there is hardly any sensitivity to the Higgs boson mass below 115\,GeV, although this region is already excluded by the LEP experiments.
 
The accuracy of the Higgs boson mass determination is estimated in the same way as in the VBF analysis. Table~\ref{tab:StatError_ggF} lists estimates of the statistical errors, $\Delta^{stat.}m_H$, defined by eq.~\eqref{eq:StatError}. (See comments on table~\ref{tab:StatError_VBF_1LepTrig} for detailed definitions of the values in table~\ref{tab:StatError_ggF}.\,) The upper and lower rows list the upper and lower bounds of the errors, respectively. We cannot evaluate the lower bounds of the errors, reflecting the closeness of $I(m)$ with different $m_H^{\text{MC}}$. Despite much higher statistics of the ggF events compared to the VBF case, the statistical errors for the ggF process do not get better, due to its background domination. We note that, since we use the leading-order cross section, these estimates may be conservative. We expect that this background domination should cause the same problem for any other methods for the determination of the Higgs boson mass using the same process.

\begin{table}[tbp]
	\centering
	\begin{tabular}{|ccc|c|}
		\hline 
		~~~~~$ee$~~~&~~~$e\mu$~~~
		&~~~$\mu\mu$~~~~&~~Combined~~\\
		\hline 
		   $+39$ & $+28$ & $-$ & $+23$\\ 
		  $-$ & $-$ & $-$ & $-$\\ 
		\hline
	\end{tabular}
	\caption{\label{tab:StatError_ggF} The estimated statistical errors (GeV), $\Delta^{stat.} m_H$, for the Higgs mass reconstruction with an integrated luminosity of 40\,fb$^{-1}$. The weight function given in eq.~\eqref{eq:WeightFuncForVBFHiggs} with $n=4$ is used. The upper and lower rows list the upper and lower bounds of errors, respectively. `$-$' denotes that we cannot evaluate the error.}
\end{table}

On the other hand, systematic uncertainties for the ggF process are well suppressed compared to statistical errors. We estimate the uncertainties of $m_H$ from the jet energy scale as $2\%$, from the factorization scale for the signal events as $5\%$, and from the $WW$ background normalization as below $1\%$ by the same method as in the VBF analysis.

\section{Conclusions}
\label{sec:Conclusion}

Accurate measurements to reveal properties of the Higgs boson and other possible new particles are essential to understanding physics behind them. In order to overcome difficulties associated with hadron collider experiments, we have introduced theoretically new quantities, characteristic weight functions. We have found an infinite number of weight functions with the following characteristics. For a many-body decay, $X \rightarrow \,l \,+\, \text{anything}$, where $X$ is a scalar or unpolarized particle and $l$ is lepton $e$ or $\mu$, the integral of the lepton energy distribution weighted by such a weight function is zero, irrespective of the velocity distribution of $X$. Using these weight functions, we have proposed a new method to measure various physical parameters even for processes with missing momenta in the final state. We call it the weight function method. This method requires only the lepton energy distribution obtained by an experiment, and ideally we do not need to know the velocity distribution of the parent particle, that is, this method does not suffer from uncertainties in the production process of the parent particle. In real experiments, however, there are many factors we should take into account, such as event selection cuts and backgrounds, and the method becomes more involved.

We have applied the weight function method to a reconstruction of the Higgs boson mass in the $H \rightarrow WW \rightarrow l\nu l\nu$ decay process at the LHC. We have performed a Monte Carlo simulation analysis including cuts and main background effects, assuming the true Higgs mass to be $m_H=125$\,GeV. In vector-boson fusion production (the VBF channel), we have estimated the statistical accuracy of the mass determination with the weight function method to be $+12\%$ and $-14\%$ at $\sqrt{s}=14$\,TeV corresponding to an integrated luminosity of 100\,fb$^{-1}$. We have found that systematic errors are suppressed, compared to the statistical errors. We note that these results may be conservative estimates since the cross sections would be underestimated in this analysis. We have found that the accuracy of mass determination depends strongly on $m_H$ and it improves to $10\%$ for $m_H=130$\,GeV and reaches $2\%$ for $m_H=150$\,GeV~\cite{Kawabata:2011gz}. It has been shown that the weight function method is sensitive to the cuts concerning leptons, whereas it is stable against cuts concerning jets. Especially, the lepton $p_T$ trigger, which strongly deforms the low energy part of the lepton distribution, worsen the sensitivity to $m_H$. We can reduce this effect by using di-lepton triggers instead of single lepton triggers. In gluon-gluon fusion production (the ggF channel), we have analyzed $H+0$-jet channel at $\sqrt{s}=8$\,TeV. We have found that even though the statistics of the ggF events are much larger than the VBF case, it is difficult to determine the mass accurately in the ggF case due to its background domination. Also in this channel, systematic errors are found to be suppressed compared to the statistical errors.

To summarize, by using the weight function method, ideally, we can avoid two major sources of uncertainties in an accurate measurement to uncover properties of a particle in hadron collider experiments, namely, uncertainties associated with jets in the final states and uncertainties in the velocity distribution of the particle. In real experiments these ideal features are affected by cuts and acceptance corrections and by contributions from background events. We find (in the case of a Higgs mass reconstruction) that the major effects of the former stem from lepton $p_T$ cuts. Thus, we expect that these effects can be predicted accurately using MC simulations. On the other hand, the latter effects need to be understood accurately experimentally, for instance, with a side-band method.

For today's realistic value of the Higgs boson mass, it is challenging to perform an accurate measurement of $m_H$ via the VBF and $H \rightarrow WW \rightarrow l\nu l\nu$ channel. Due to small statistics, the statistical error will dominate over the systematic one, so that the advantage of the weight function method cannot be utilized in an optimal way. In particular, a part of the whole signal events are effectively not used in our method, due to a projection by the weight function (the part dependent on the velocity of the Higgs boson do not contribute). Hence, this method is disadvantageous in terms of the statistical error, although there is some room for adjustment using the degree of freedom of the weight function. A naive estimate indicates that fitting the whole lepton energy spectrum  for a reconstruction of the Higgs boson mass can give a few tens $\%$ better statistical error. Thus, the present status of our study stays to be somewhat demonstrative of how the method works in principle (which is seen clearly in the case study for $m_H=150$\,GeV~\cite{Kawabata:2011gz}). We also note that our result serves as a reference point, since up to now there are only few studies on the Higgs mass reconstruction using this channel for $m_H \approx 125$\,GeV.

We expect that the weight function method has wide applications. Above all, a mass reconstruction of the top quark will be an attractive work. In addition, applications to particles appearing in models beyond the Standard Model, such as scalars in an extended Higgs sector, can be interesting. We mention that the Higgs mass reconstruction using $H \rightarrow \tau \tau$ decay mode may also be possible although it would be challenging to avoid effects by lepton $p_T$ cuts.

\appendix
\section{Lepton energy distribution with a constraint $M_{ll}<M$}
\label{sec:DerivationOfLepEneDist}

In this appendix we give a derivation of~\eqref{eq:LepEneDistWithMll}, the lepton energy distribution for the $H\rightarrow W^+W^- \rightarrow l^+ \nu l^-\overline{\nu}$ process with the requirement for the invariant mass of the leptons $M_{ll}<M$ imposed. 

The amplitude for this process is given by
\begin{eqnarray}
	\mathcal{M} &=& \frac{g^3 M_W}{8} 
					\left[ \,\overline{u_\nu}\gamma_\alpha (1-\gamma_5)v_{l^+} \right] 
					\frac{1}{p_{W^+}^2-M_W^2+iM_W\Gamma_W} 
					\left(g^{\alpha \beta}-\frac{p_{W^+}^{\alpha}p_{W^-}^{\beta}}{M_W^2}\right) 
					g_{\beta \delta} \nonumber \\
			    &   & ~~~~~~\times \left[\, \overline{u_{l^-}} \gamma_\lambda (1-\gamma_5)v_{\overline{\nu}} \right] 
			    		\frac{1}{p_{W^-}^2-M_W^2+iM_W \Gamma_W} 
					\left( g^{\lambda \delta}- \frac{p_{W^-}^\lambda p_{W^-}^\delta}{M_W^2} \right) \,,
\end{eqnarray}
at tree level. Neglecting the masses of the leptons, one obtains
\begin{eqnarray}
	\Sigma \, |\mathcal{M}|^2 = g^6 M_W^2~ \frac{1}{\left(p_{W^+}^2-M_W^2 \right)^2+M_W^2 \Gamma_W^2} 
									\,\frac{1}{\left(p_{W^-}^2-M_W^2\right)^2+M_W^2\Gamma_W^2}
									\,4\,(p_{l^+}\cdot p_{\overline{\nu}})(p_\nu \cdot p_{l^-})\,,\nonumber \\
\end{eqnarray}
where $\Sigma$ denotes the sum over final spins. Imposing $M_{ll}<M$, the decay rate becomes
\begin{eqnarray}
	d\Gamma_{M_{ll}<M} &=& \frac{4g^6M_W^2}{2p_H^0} \int\!\!\!\int_0^\infty \frac{d\mu^2}{2\pi} \frac{d\overline{\mu}^2}{2\pi}\,
			 	\frac{1}{(\mu^2-M_W^2)^2+M_W^2\Gamma_W^2}\, 
				\frac{1}{(\overline{\mu}^2-M_W^2)^2+M_W^2\Gamma_W^2}\nonumber \\[1ex]
			 &  & ~\times (p_{l^+}\cdot p_{\overline{\nu}})(p_\nu \cdot p_{l^-})\nonumber \\[1ex]
			 &  & ~\times d\Phi_2(H \rightarrow W^{+}W^{-}) d\Phi_2(W^{+} \rightarrow l^+ \nu) 
			 	d\Phi_2(W^{-} \rightarrow l^- \overline{\nu})\nonumber \\[1ex]
			 &  & ~\times \theta (M-M_{ll})\,,
	\label{eq:DecayRateWithMll}
\end{eqnarray}
where $d\Phi_2$ is the phase space for the two-body decay:
\begin{eqnarray}
	d\Phi_2(X \rightarrow YZ) &\equiv& (2\pi)^4 \delta^4(p_Y+p_Z-p_X) \frac{d^3\bm{p}_Y}{(2\pi)^3 2p_Y^0} \frac{d^3 \bm{p}_Z}{(2\pi)^3 2p_Z^0}\\
	&&\!\!\!\!\!\!\!\!\!\mu^2 \equiv p_{W^{+}}^2\,,~ ~{\overline{\mu}}^2 \equiv p_{W^{-}}^2\,.
\end{eqnarray}

We carry out the integrations for the phase space. After integrating over $p_{\overline{\nu}}$, $p_\nu$ and $p_{W^+}$:
\begin{eqnarray}
	\begin{aligned}
		&(p_{l^+}\cdot p_{\overline{\nu}})(p_\nu \cdot p_{l^-})
			\,d\Phi_2(H \rightarrow W^{+}W^{-}) \,d\Phi_2(W^{+} \rightarrow l^+ \nu)
			\,d\Phi_2(W^{-} \rightarrow l^- \overline{\nu}) \\[1ex]
		&~~=~ \frac{1}{(2\pi)^6} \left\{p_{l^+} \cdot (p_{W^-}-p_{l^-})\right\} (p_\nu \cdot p_{l^-}) \\[-0.5ex]
		&~~~~~~~\times \delta \!\left( (p_H-p_{W^-})^2-\mu^2\right) \theta \!\left( p_H^0-p_{W^-}^0\right) 
			\frac{d^3 \bm{p}_{W^-}}{2p_{W^-}^0} \\[-0.5ex]
		&~~~~~~~\times \delta \!\left( (p_H-p_{W^-}-p_{l^+})^2\right) \theta \!\left( p_H^0-p_{W^-}^0-p_{l^+}^0\right) 
			\frac{d^3 \bm{p}_{l^+}}{2p_{l^+}^0} \\[-0.5ex]
		&~~~~~~~\times \delta \!\left( (p_{W^-}-p_{l^-})^2\right) \theta \!\left( p_{W^-}^0-p_{l^-}^0\right) 
			\frac{d^3 \bm{p}_{l^-}}{2p_{l^-}^0}\,,
	\end{aligned}
\end{eqnarray}
we perform the remaining integrations in the following order:
\begin{equation}
	\bm{p}_{l^-} ~\rightarrow ~~\bm{p}_{W^-} ~\rightarrow ~~\bm{p}_{l^+} ~\rightarrow ~~\overline{\mu}^2 ~\rightarrow ~~\mu^2
\end{equation}

First, we integrate over $\bm{p}_{l^-}$ in the rest frame of $W^-$. The condition $M_{ll}<M$ implies
\begin{equation}
	\cos \tilde{\theta}\, >\, 1-\,\frac{M^2}{2 \tilde{E}_{l^+} \tilde{E}_{l^-}}\, \equiv ~z\,,
\end{equation}
where a variable with a tilde denotes a quantity defined in the rest frame of $W^-$, and $\tilde{\theta}$ is the polar angle between the directions of $l^+$ and $l^-$. Hence, the integral region of $\cos \tilde{\theta}$ is given by
\begin{eqnarray}
	\begin{aligned}
		\begin{cases}
			~-1 &\!\!< ~\cos \tilde{\theta} ~<~1\hspace{1.5cm} {\rm for} ~~~E_{l^+} < (M^2+\mu^2)/2m_H\\
			~~~~z &\!\!< ~\cos \tilde{\theta} ~<~1\hspace{1.5cm}{\rm for} ~~~E_{l^+} > (M^2+\mu^2)/2m_H\,.
		\end{cases}
	\end{aligned}
\end{eqnarray}
One finds the integral over $\tilde{\theta}$ and $\tilde{\phi}$
\begin{eqnarray}
	\begin{aligned}
		~~I(x) ~&\equiv~ \frac{1}{2\pi} \int_x^1 d\cos \tilde{\theta} \int_0^{2\pi} d\tilde{\phi} 
			\left\{ p_{l^+} \cdot \left( p_{W^-}-p_{l^-}\right)\right\} \left( p_{\nu} \cdot p_{l^-}\right)\\[1ex]
		&=~\tilde{E}_{l^+} \tilde{E}_{\nu}\, \tilde{E}_{l^-}
			\left[ (\overline{\mu}-\tilde{E}_{l^-})(1-x)
				+\left\{ \tilde{E}_{l^-} -(\overline{\mu}-\tilde{E}_{l^-}) \cos \tilde{\theta}_\nu\right\} \frac{1-x^2}{2}
				-\tilde{E}_{l^-} \cos \tilde{\theta}_\nu \frac{1-x^3}{3}\,\right]\,,
	\end{aligned}\nonumber\\[0.5em]\label{eq:I(x)}
\end{eqnarray}
where $x=-1$ or $z$ and $\tilde{\theta}_\nu$ is the polar angle between the directions of $l^+$ and $\nu$, and the integral over $\tilde{E}_{l^-}$ becomes
\begin{eqnarray}
	\int \,\delta \!\left( (p_{W^-}-p_{l^-})^2\right) \theta \!\left( p_{W^-}^0-p_{l^-}^0\right) \frac{\tilde{E}_{l^-}}{2} \,d\tilde{E}_{l^-}
		~=~ \frac{1}{8}\,.
\end{eqnarray}

The integrations over $\bm{p}_{W^-}$ and $\bm{p}_{l^+}$ are carried out in the rest frame of the Higgs boson and in the rest frame of $W^+$, respectively. The former integral over the solid angle just gives $\int d\Omega =4\pi$. The latter gives
\begin{equation}
	\delta \!\left( (p_{W^+}-p_{l^+})^2\right) \theta \!\left( p_{W^+}^0-p_{l^+}^0\right) 
			\frac{d^3 \bm{p}_{l^+}}{2p_{l^+}^0}
	\,=\, \frac{\pi}{2 \left| \bm{p}_{W^-}\right|}~ dE_{l^+} \,\theta \,(\,E_{lmin} < E_{l^+} < E_{lmax}\,)
\end{equation}
with
\begin{equation}
	E_{l_{min}^{max}} ~\equiv~ \frac{1}{2} \left(\,E_{W^+} \pm \,\left| \bm{p}_{W^+}\right| \,\right)\,.
\end{equation}

Next, we integrate over $\overline{\mu}^2$:
\begin{eqnarray}
	\begin{aligned}
		J(x) &\equiv \,\int_0^\infty d\overline{\mu}^2\,\frac{1}{(\overline{\mu}^2-M_W^2)^2+M_W^2\Gamma_W^2}
			~I(x) \,\theta (E_{W^+}) ~\theta (\,E_{lmin} < E_{l^+} < E_{lmax}\,)\\[1ex]
			&= \,\int_0^{\overline{\mu}_{max}^2} d\overline{\mu}^2
			\,\frac{1}{(\overline{\mu}^2-M_W^2)^2+M_W^2\Gamma_W^2} ~I(x)\\[1ex]
			&= \,\frac{i}{2M_W \Gamma_W}  \left[ A_+(x) \log \,({\overline{\mu}}^2_{max}-M_W^2+iM_W \Gamma_W)
					-A_+(x)\log \,(-M_W^2 +iM_W \Gamma_W)\right. \nonumber \\[0.5ex]
			&~~~~~~~~~~~~~~~~\left. -A_-(x) \log\,(\overline{\mu}^2_{max} -M_W^2 -iM_W \Gamma_W)
					+A_-(x) \log\, (-M_W^2 -iM_W \Gamma_W) \right]\,.
	\end{aligned}
\end{eqnarray}
Here, $\overline{\mu}_{max}^2$ and $A_\pm(x)$ are defined by
\begin{equation}
	\overline{\mu}_{max}^2 ~\equiv~ m_H^2 + \mu^2 - \frac{m_H (4E_{l^+}^2 + \mu^2)}{2E_{l^+}} \,,
\end{equation}
and
\begin{eqnarray}
	A_\pm(x) &\equiv& -\left[ \frac{\mu^2}{8}\left( \frac{1-x^2}{2} +\frac{1-x^3}{3}\right)
					-\frac{1}{16} \left( 2m_H E_{l^+}-\mu^2\right) \left( 1-x-\frac{1-x^3}{3} \right)\right]
					(-M_W^2\pm iM_W\Gamma_W)\nonumber \\
			&          &  ~~+\frac{1}{16} (2m_H E_{l^+}-\mu^2)(m_H^2-2m_H E_{l^+})\left( 1-x-\frac{1-x^3}{3}\right)\,.
\end{eqnarray}

As a result, we complete the integrations of the phase space except over $\mu^2$, and obtain the expression~\eqref{eq:LepEneDistWithMll}.

\acknowledgments

The works of S.K. and Y.S., respectively, were supported by Grant-in-Aid for JSPS Fellows under the program number 24$\cdot$3439 and by the Japanese Ministry of Education, Culture, Sports, Science and Technology by Grant-in-Aid for Scientific Research under the program number (C) 23540281. The work of H.Y. was supported in part by Grant-in-Aid for Scientific Research, No.\,24340036 and the Sasakawa Scientific Research Grant from The Japan Science Society.


\end{document}